\documentclass[useAMS, usenatbib]{mn2e}
\usepackage{graphicx} % This allows you to call \includegraphics
\usepackage{rotating}
\usepackage{subfigure}
\usepackage{multirow}
\usepackage{verbatim}

\usepackage{amssymb}
\def\eq{\begin{equation}}
\def\en{\end{equation}}

\def\P3hat{{\mathaccent 94 P}_3}

\newcommand{\simgt}%
        {\,\hbox{\lower0.6ex\hbox{$\sim$}\llap{\raise0.6ex\hbox{$>$}}}\,}
\newcommand{\simlt}%
        {\,\hbox{\lower0.6ex\hbox{$\sim$}\llap{\raise0.6ex\hbox{$<$}}}\,}
\def\gap{\hbox{\hspace{1.5mm}}\raise2pt
       \vbox{\hbox{$>$}}\lower2pt
       \vbox{\moveleft7.0pt\hbox{$\sim$ }}\hbox{\hskip 0.05mm}}
\def \mref#1{(\ref{#1})}

\newcommand\sss{\scriptscriptstyle}

\def \dcut{\delta_{\rm cut}}
\def\thm{\theta_{m}}
\def\phm{\phi_{m}}
\def\rlc{R_{lc}}
\def\dsp{\Delta^\theta_{\rm sp}}
\def\dspl{\Delta^{\sss \negthinspace L}}
\def\dspt{\Delta^{\sss \negthinspace T}}
\def\dspp{\Delta^{\rm \negthinspace pair}}
\def\dspc{\Delta^{\rm \negthinspace ctc}}
\usepackage{rotating} % to rotate the expressions and tables.
\title[Frequency evolution of pulsar radio profiles]
{The origin of the frequency-dependent
behaviour \\ of pulsar radio profiles}

\author[J.~Dyks \& Rudak] 
{J.~Dyks \& B.~Rudak\\
Nicolaus Copernicus Astronomical Center, Rabia\'nska 8, 87-100, Toru\'n,
Poland\thanks{jinx@ncac.torun.pl} }

\date{Originally drafted 2014 07 07}

\pagerange{\pageref{firstpage}--\pageref{lastpage}} \pubyear{2011}

\def\LaTeX{L\kern-.36em\raise.3ex\hbox{a}\kern-.15em    T\kern-.1667em\lower.7ex\hbox{E}\kern-.125emX}

\begin{document}

\label{firstpage}

\maketitle

\begin{abstract}
We present further development of a pulsar emission model 
based on multiple streams
diverging away from the magnetic dipole axis, 
and forming azimuthally-structured 
fan-shaped beams. It is shown that this geometry, successfully tested on
profiles with bifurcated features, naturally solves
several classical pulsar problems and avoids some difficulties
of the traditional nested cone/core model. This is best visible for profiles
with several components, such as those of class T, Q and M, because they
most clearly exhibit a range of effects previously interpreted within the
conal model. In particular, with no reference to the flaring boundary
of the polar magnetic flux tube, the stream model
explains the apparent radius-to-frequency mapping (RFM), including
its reduced strength for the inner pair of components.
The lag of the central component (apparent `core') with respect to the 
centroids of the flanking (`conal') components
can also be naturally explained with no reference to emission rings
located at disparate altitudes. The stream model also 
reveals why the millisecond pulsars, despite their more strongly flaring
magnetic field lines, do not exhibit as strong RFM as the normal
pulsars. 
The model is then successful 
in reproducing properties of
so disparate objects as the M-class and millisecond pulsars,
including some peculiarities of the latter.
With no hesitation we, therefore, advance the view that pulsars
have fan beams generated by outflowing streams, whereas
the nested cone/core beams may well not exist at all.
\end{abstract} 

\begin{keywords}
pulsars: general -- pulsars: individual: B0329+54 -- Radiation mechanisms:
non-thermal.
\end{keywords}

\maketitle

\section{Introduction}
\label{sectI}
The first mention of streams we are aware of was that of 
Michel (1987).
The main observational evidence on which he based his idea 
was the sign reversal of circular polarisation at peaks
of components in radio profiles. However, a detectable sign reversal 
could also be expected for other geometries 
of the beam, 
eg.~the patchy beam (Lyne \& Manchester 1988; Manchester 2012)
and the conal one (Rankin 1983), especially if the latter is structured
as the `patchy conal ring'
in Fig.~1 of Karastergiou \& Johnston (2007).
A carefull recent study of the sign reversing circular polarisation 
is presented in Gangadhara (2010), Wang et al.~(2012), and Wang et
al.~(2010).
Moreover, Michel's fan beams were very elongated
($\Delta\theta \sim 1$ rad), and more complicated than the single-parameter
conal beams, which seemed to be so well designed to match the imposing
flaring polar tube. They were also less attractive, with little inherent symmetry
which could automatically explain the approximate left-right symmetry 
of some pulsar profiles (notably of Q and M class in the classification 
scheme of Rankin 1983). This may be why Michel's model was not 
recognised as a promising way to interpet radio pulsar profiles
and have in principle been ignored afterwards.

A major development of the stream-based (or fan beam) model
has recently been sparked by the properties of bifurcated features residing
in pulsar profiles
(Dyks et al.~2010a, hereafter DRD10; 
Dyks \& Rudak 2012, hereafter DR12). DRD10
have shown that double notches (McLaughlin \& Rankin 2004; Navarro et
al.~1997; Rankin \& Rathnasree 1997)
can be understood in a natural way, if the elementary emitting unit
has the form of an elongated stream emitting a bifurcated
beam. This model required that both emitting streams
and obscuring streams exist in pulsar magnetosphere.
The former ones were immediately identified as the \emph{apparently} conal
components in the profile of PSR J0437$-$4715, and the bifurcated precursors
in PSR J1012$+$5307. The bifurcation of a seemingly `conal' component
in PSR J0437$-$4715, and the fixed (frequency-independent) location of
the `conal' components and notches have suggested that all these features
are produced by a cut of sightline through narrow fan beams
emitted by azimuthally-narrow streams. 

A change of the peak-to-peak separation between components observed at
different frequency $\nu$, and the associated change of pulse profile width,
can be interpreted in terms of the radius-to-frequency mapping
(RFM, Komesaroff 1970; Cordes 1978).
According to the RFM model, the low-$\nu$ radiation is emitted at a larger
radial distance $r$, measured from the neutron star centre, than the
high-$\nu$ radiation. However, the RFM acronym tends to be simultaneously
used for the observed effect itself, ie. for the $\nu$-dependent
displacement of components.
To discern between the observed phenomenon and the geometrical
interpretation, for the former one we will hereafter use the name 
{\it apparent
RFM}. To avoid multiple adjectives, sometimes we will just write `RFM'
(in single quotation marks) in the same sense. Consequently, `RFM' refers to the
observed change of components' separation, whereas RFM to the geometrical
model (spatial association of $\nu$ with $r$).

The $\nu$-independent  separation of `conal' components in pulsar profiles,
ie.~the lack of apparent radius-to-frequency mapping (`RFM') 
may indeed be considered as a direct
consequence of the stream-cut geometry: 
in the limit of an infinitely narrow stream, any radiation
(of arbitrary frequency) can only be observed 
when the line of sight is in the plane of the stream.
Narrow streams then provide excellent conditions
for the lack of the apparent RFM.
The profile of PSR J0437$-$4715, with its fixed location of components, 
bifurcations, and double notches has then been identified 
as a clear example of a profile in which the seemingly conal components
are in fact not conal at all 
-- they originate from the sightline passage through streams
which, when viewed down the dipole axis, resemble a star-shaped
pattern, or a pattern created by blades of an indoor ceiling fan.

Double notches require the obscuring streams or stream-shaped
non-emitting regions within a radially- and laterally-extended emitter.
Since the notches are also observed in normal pulsars (B1929$+$10, and
B0950$+$08), DRD10 have concluded that streams must be
ubiquitous among pulsars. Therefore, the need to extend 
the conal classification scheme
to include the stream-shaped emitters, has been emphasized
 (see Fig.~18 in DRD10).

Precessing pulsars provide
a direct way to map pulsar beams in 2D.
However, conal beam models have not been able to reproduce
the observed pulse evolution precisely
(Kramer 1998; Clifton \& Weisberg 2008; Lorimer et al.~2006; Burgay et al.~2005).
This has lead to proposals of several beam shapes 
(hourglass, Weisberg \& Taylor 2002; horseshoe, Perera et al.~2010) 
which nevertheless had overall morphology similar to the conal beam.
DRD10 (Sect.~6.3.3 therein) suggested that 
the problems with the modelling of precessing beams 
are caused by the use of conal beams instead of the fan-shaped ones.
A fan-shaped beam has explicitly been proposed for J1906$+$0746,
based on the stability of its main pulse shape on a few-years-long timescale.

That prediction of a stream emitter in J1906$+$0746 
has recently been confirmed by the beam map of Desvignes et al.~(2013),
although their geometrical solution is not well constrained
(dipole tilt $\alpha$ with respect to the spin axis 
is equal to $81^{+1}_{-66}$ degrees).
An azimuthally-limited emitter has also been
deduced by Manchester et al.~(2010).

The maps of precessing pulsars have recently been reproduced in 
Wang et al.~(2014),  
who analyse statistical distributions of average profile properties in
terms of both conal and fan models. Their distribution of pulse widths as a
function of impact angle provides further credence for the ubiquity of the
fan-shaped geometry of pulsar beams.

A major obstacle for the acceptance of the fan-beam geometry
comes from the multiple-component profiles classified as T-, Q- and M-type by
Rankin (1983; 1993). This is because they exhibit a wide range of phenomena
that look deceptively conal. In particular, the peak-to-peak separation
between their components increases with decreasing frequency, and
this phenomenon is stronger for the outermost pair (eg.~Mitra \& Rankin
2002). For years this has been interpreted as the consequence of the fact
that in dipolar magnetic field the outer circumpolar field lines 
are more strongly flaring away from the dipole axis than the inner ones.

The conal model was also applied to 
the lag of the central component  with respect to
centroids of `conal' pairs, as observed in B0329$+$54 
(Gangadhara \& Gupta 2001, hereafter
GG01) and other objects (Gupta \& Gangadhara 2003, hereafter GG03).
However, the lag had to be interpreted through
a complicated structure of a near-surface emission region and several
 emission rings  located at disparate
altitudes. Altitude-dependent effects of aberration and retardation (AR), 
differentially deflecting the emission from the rings, 
were responsible for the observed delay. 
The AR effects with spatial RFM were also able
to justify why the magnitude of the lag increases 
for decreasing frequency 
(GG01; Dyks et al.~2004).

The conal beam model was also supported by
the statistical distributions of pulse width (and components' separation)
versus rotation period
(Rankin 1990; 1993; Gil et al.~1993; Kramer et al.~1994).
All of them seem
to suggest that widths of pulsar profile (and angular radii of cones)
follow the inverse square root
dependence on the rotation period $P$. 
This could imply that the fixed-intensity outer envelope 
of pulsar emission and dimensions of cones
follow the flaring shape of polar tube. 
The derived dimensions of cones could even find a theoretical justification
(Wright 2003).
However, this does not exclude the possibility of 
the azimuthal structure of the beam.
Moreover, the axisymmetric beam geometry
is involved in some assumptions 
(eg.~determination of $\alpha$ from the width of the central component)
that have been used to obtain these results.

In this paper we apply the implications of the broad-band emitting
streams to normal radio pulsars, especially those of T, M and Q type,
which can probably be considered as the prototype of conal thinking.  
We focus mostly on the frequency evolution of profiles, 
using the approach of DR12 (Fig.~2 therein), 
ie.~by considering the geometry of 
fixed-intensity patterns in the vicinity of the line of sight.  
Several results of this paper apply also for the patchy beams of Lyne \&
Manchester (1988), since a
single fan-shaped subbeam can be considered as a type of an elongated
patch (Fig.~\ref{params}). We will also consider
circular patches of emission as a
zeroth-order model for more realistic elongated beams.
However, in the case of a fan beam emitted by a
stream, the beam's elongation is explicitly identified: it is latitudinal
and related to the curvature of B-field lines. 
In the case of a general patchy beam, the elongation or geometry of patches
is less constrained.
For example, if a patch is elongated mostly in the magnetic azimuth,
then the location of components in a profile is determined by the conal
aspects
of geometry. This is the case for the partial cones shown in Fig.~18 of
DRD10 (middle sightline). In the present paper we focus on the
latitudinally-extended patches (fan beams).

The paper is organized as follows. 
After a short introduction of key parameters of our model
(Sect.~\ref{PARAMS}) we show how the model leads to the apparent 
RFM effect (Sect.~\ref{RFM}). In Sect.~\ref{mapping} we 
connect the apparent RFM with the spatial emissivity distribution 
in pulsar magnetosphere. The lag of a central component with respect
to a midpoint between outer components is explained in
Sect.~\ref{ar}, followed by Sect.~\ref{obscur} which discusses
possible  obscuration effects.

\section{Main parameters of the stream model}
\label{PARAMS}

\begin{figure}
	\centering
	\includegraphics[width=0.48\textwidth,]{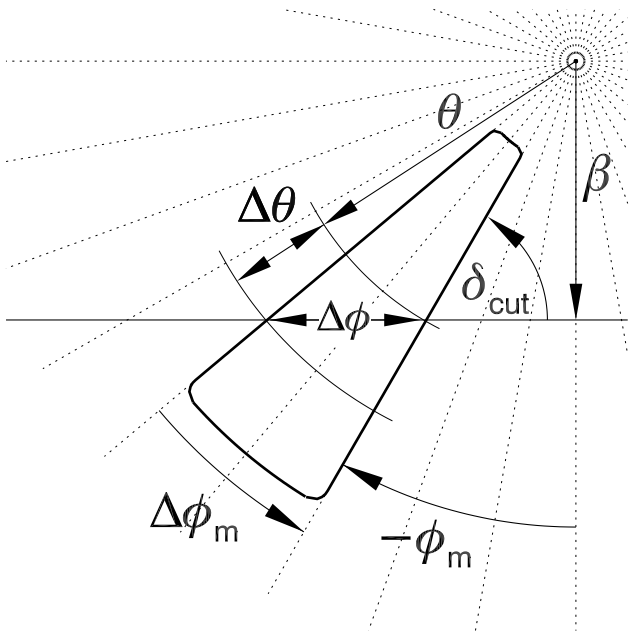}
\caption{Sky-projected geometry of azimuthally-limited emission 
beam (thick contour) viewed down the dipole axis (top right dot). 
The horizontal line marks the path of the line of sight. 
Dotted lines present dipolar magnetic field lines.
The azimuth $\phm$ is measured counterclockwise from the main meridional plane.}
\label{params}
\end{figure}

The dipolar magnetic field is assumed in this work.
Because of the divergence of dipolar field lines, the radio-emitting region,
when viewed down the dipole axis $\vec \mu$, typically has
the form of a wedge. The region extends over a limited range of magnetic
azimuth $\phm$ measured around the dipole axis, as well as over
 a limited range of magnetic colatitude $\thm$, which is the angle
 between the radial position vector of an emission point and the dipole axis.
In the dipolar field, radiation from a point at $\thm$ propagates at the
angle $\theta\approx(3/2)\thm$ with respect to $\vec \mu$.
Therefore, the associated beam (thick-line contour in Fig.~\ref{params})
extends over a range of $\theta$ that can be directly translated into $\thm$.
The propagation angle $\theta$ is marked in Fig.~\ref{params}. 
The beam extends azimuthally over $\Delta\phm$ -- 
the same range of $\phm$ as the underlying stream.  
The latitudinal elongation of the wedge-shaped beam in Fig.~\ref{params}
results from the 2D projection of the curved stream on the sky.

\begin{figure*}
	\centering
	\includegraphics[width=0.9\textwidth]{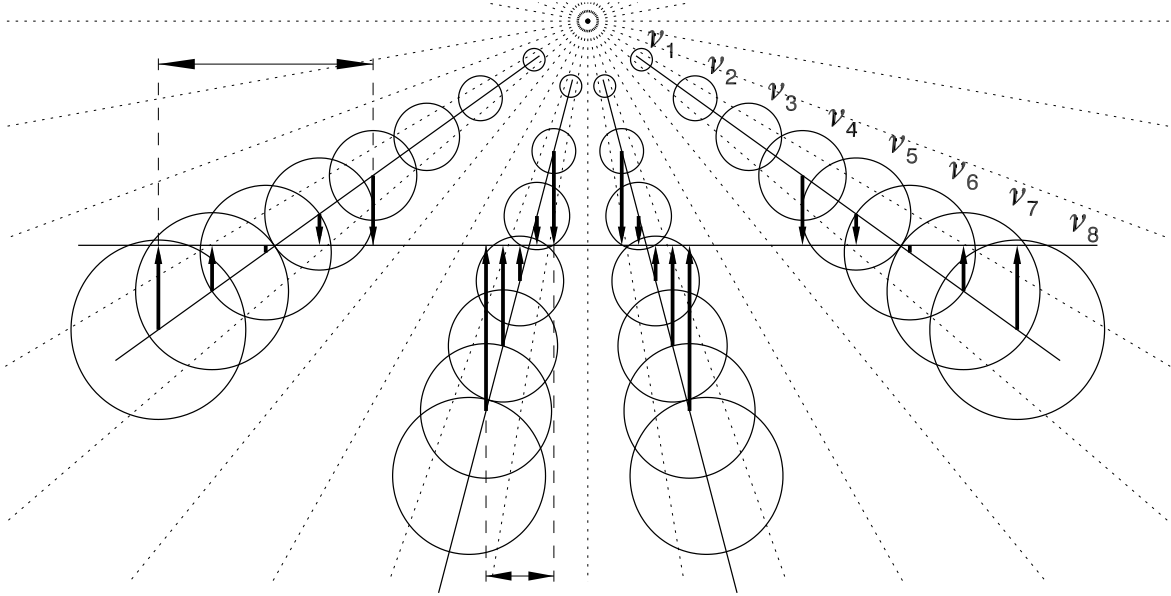}
\caption{The mechanism of the apparent radius-to-frequency 
mapping. Circles of different size represent fixed-intensity contours at
different frequencies, with the maximum intensity at the their centres. 
Longitudes of observed peak flux at different $\nu$ are marked
with vertical arrows. The equidistant locations of the consecutive 
$\nu$-contours (same for all beams) represent the same angular 
spectral gradient along all four beams. 
Note the smaller frequency-dependent spread of the inner components (bottom
horizontal arrow) compared to the outer components (top horizontal arrow).
}
\label{rfm}
\end{figure*}

The observer's sightline probes the beam along 
the horizontal\footnote{For simplicty we ignore all spherical trigonometry
issues that are not important for the illustration of our arguments.}
line in Fig.~\ref{params} and it
crosses the beam at the cut angle $\dcut$ measured between some 
fixed magnetic azimuth $\phm$ within the beam 
and the sightline's path (both of them projected on the sky). 
The passage through the wedge beam
lasts for a limited interval of pulse phase $\phi$ (pulse
longitude). 
Fig.~\ref{params} shows that the passage through the beam (lasting over 
the interval $\Delta\phi$) corresponds to a change in angular distance
from the dipole axis $\Delta\theta$. 
%For streams that are narrow in the direction of magnetic
%colatitude $\thm$, 
Assuming that streams follow the bent B-field lines, 
the passage must also be associated with a gradual change 
in detectable radial distance $r$  
from the centre of the star.
It is important that the probed range of angular distances from the dipole
axis ($\Delta\theta$) increases when the beam is cut at a smaller angle
$\dcut$.
For orthogonal cuts ($\dcut \sim 90^\circ$), as well as for extremely narrow
streams, $\Delta\theta\sim0$.

The cut angle is a crucial
parameter in the stream model. As will be shown below, it is responsible
for several aspects of the longitude-and-frequency-dependent behaviour 
established for multiple-component pulsars. In some cases
the cut angle $\dcut$ basically mimics the effect of flaring 
of polar tube. The cut angle is equal to $\dcut = 90^\circ
-\psi_{\sss PA}$ (see eq.~6 in DR12), where $\psi_{\sss PA}$ is 
the polarisation position angle measured from the main meridional plane,
containing the dipole and rotation axes.  
 In the flat case with 
large $\alpha\sim90^\circ$, we have $\dcut=90^\circ-|\phm|$
(Fig.~\ref{params}).

We additionally use the standard geometrical pulsar parameters:
 the dipole inclination angle $\alpha$ with 
respect to the rotation axis $\vec \Omega$,
the viewing angle $\zeta$ between $\vec \Omega$ and the observer's line of sight,
and the impact angle $\beta=\zeta-\alpha$. We will also use the 
footprint parameter $s$, 
which is the transverse distance from the dipole axis,
 normalized by the transverse radius of the open field line region
(both measured at the same radial distance $r$).

\section{Origin of the observed radius- -to-frequency mapping}
\label{RFM}

By the `observed RFM' (or apparent RFM, or just `RFM' in single 
quotation marks) 
we mean the change of peak-to-peak separations
between components observed at different frequencies, and the associated
change of pulse profile width.

To illustrate its origin within the stream model we will temporarily assume 
that the sky-projected 
pattern of radio emission at a fixed frequency has a circular shape
with the radio intensity decreasing away from the centre 
(Fig.~\ref{rfm}). One can imagine
a two dimensional (axially-symmetrical) Gaussian projected on the sky.
 Emission at a fixed $\nu$ is not constrained 
to the interior of a circle only.
It may be thought that the circles present, say the $50$\% intensity level, 
with the  peak emissivity at the centre of each circle. 
Assuming that the spatial RFM exists and magnetic field lines flare away
from the dipole axis, the low-frequency patterns
should be located further away from the axis. 
This is shown with the sets of solid circles in Fig.~\ref{rfm}. 
The circles that correspond
to lower frequencies become larger because 
the emitting stream/wedge becomes wider at larger distance
from the dipole axis.\footnote{The diameter of the low-$\nu$ patterns 
can also be additionally
enlarged by the increasing size of microphysical beams (elementary beams
of the emitted radiation, see DR12).}
Since the intensity at a fixed frequency is monotonously decreasing 
with the distance from the centre of each circular contour, 
a component's peak 
will be observed when the line of sight is closest to the centre of a circle 
corresponding to a given frequency.
These phases are marked with vertical arrows, which show clear change
of peak location with frequency.

\begin{figure*}
	\centering
	\includegraphics[width=0.99\textwidth,]{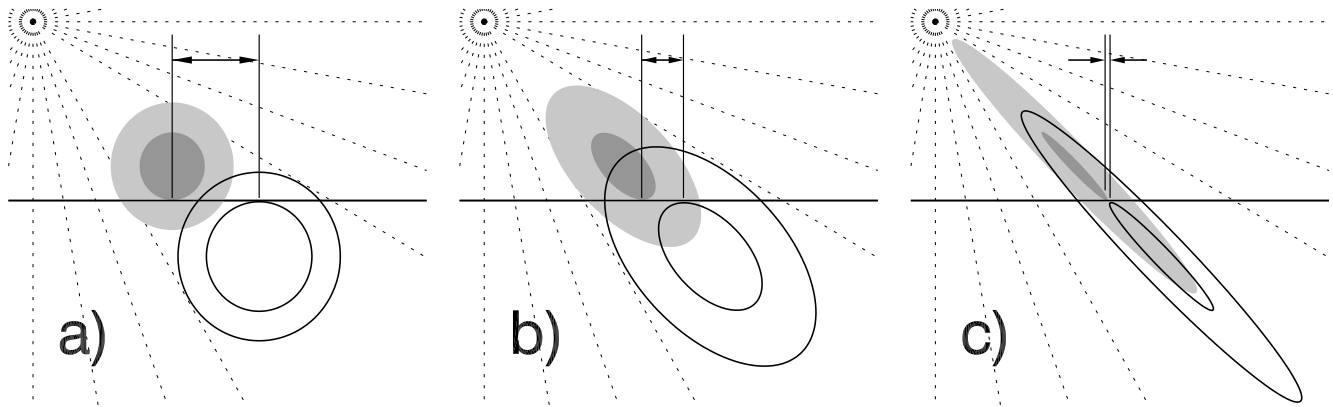}
\caption{Sky-projected emission from three streams with the same $\phm$ 
but different angular spectral structure. Fixed-intensity contours at two frequencies
($\nu_1$ and $\nu_2$) 
are shown for each beam, with two levels of intensity marked for each $\nu$. 
One frequency is marked with the grey contours,
the second -- with the solid line contours. 
The beams become more broadband from left to right,
as marked by the increasingly overlapping 
contours of different $\nu$. 
The levels of larger intensity
(inner contours) were selected to expose the location of peak flux
of observed pulse components, ie.~they are tangent to the path of the line of
sight (the solid horizontal).
The marked phase intervals present the dislocations of components
which correspond to the fixed frequency interval $\nu_1 - \nu_2$. 
{\bf a)} The circular fixed-$\nu$ contours,
% with no extent along the stream, 
which are useful to expose the role of $\dcut$. {\bf b)}
More realistic, $\theta$-extended contours for a moderately narrow-band
emission. {\bf c)} 
A broad-band beam %with a small spectral gradient in $\theta$, 
expected
for MSPs 
and pulsars with no apparent RFM.
In c), the apparent RFM is additionally decreased by 
the narrowness of the beam.
}
\label{specgrad}
\end{figure*}

An equivalent description is to say that the emitted 
spectrum changes along the stream, and, during the passage through the
stream, the line of sight samples different spectra.
However, if the angular variations of a locally-emitted spectrum
are ignored, the interpretive capabilities of the model are
seriously hampered.
Therefore, we introduce the `angular spectral gradient' which provides 
a measure
of how strongly the spectrum changes with angular separation on the
sky. 
For example, the $\theta$-component of the gradient can be defined as
$d\negthinspace\left<\nu\right>\negthinspace/d\theta$, 
where $\left<\nu(\phm,\theta)\right>$ 
is the average
frequency emitted in direction $(\phm,\theta)$, 
weighed by the spectral distribution of intensity. 
Unless otherwise specified, the term `angular spectral gradient'
will hereafter refer to the change of spectrum with $\theta$
(at a fixed $\phm$).
It will also be called the `$\theta$ spectral 
gradient'.\footnote{The $\nu$-dependent intensity can change also 
in $\phm$ direction
 (Fig.~\ref{specgrad}) so the gradient is two-dimensional. For brevity,
however, we will use the name `gradient' instead of the `$\theta$-component
of the gradient'.}
Below we will also use other types of spectral gradients, 
eg.~the radial (spatial) one.  
We will also refer to the angular scale of spectral change in direction
$\theta$. This $\theta$ spectral scale 
is proportional to $d\theta_{\rm pk}/d\nu$, where $\theta_{\rm pk}(\nu)$
is the colatitude of the centre of a pattern at the frequency $\nu$. 
 In Fig.~\ref{rfm}, this scale 
is visualised as the
separation of the circular contours corresponding to different 
frequencies.
It is worth to emphasize that the same-size contours 
on different streams  in Fig.~\ref{rfm} 
do not necessarily need to represent the same $\nu$. Their size increases just
to represent the widening of the emission region with $\theta$.

When the line of sight cuts through a few streams with similar angular 
spectral gradient, the inner components exhibit weaker `RFM'
(Fig.~\ref{rfm}). This is
because the beams of inner streams are cut more orthogonally, 
and a given angular displacement of the fixed-$\nu$ patterns,
translates to a smaller shift in pulse longitude.
This is marked in Fig.~\ref{rfm} with two horizontal arrows. Note that
the $\theta$-separations of the fixed-$\nu$ patterns 
are the same for all streams in Fig.~\ref{rfm} just to illustrate
how a given angular spectral gradient transfers onto the strength of the
observed `RFM'. In real pulsars the angular spectral gradient
may depend on $\theta$ and may
be different for streams with different $s$.
However, it may also be different for streams which differ in $\phm$ only,
because the acceleration strength can vary with the distance from
the main meridian (MM),  which contains both the rotation axis
and dipole axis
 (eg.~Arons \& Scharlemann 1979).
As we discuss in Sect.~\ref{mapping}
the spatial RFM illustrated by the circles in Fig.~\ref{rfm}
is likely caused by both the radial and transverse drop of density
in the emitting streams.

An immediate consequence of the stream geometry is the expectation of
weak (or nonexistent) `RFM' in three cases:
1) When the spectral properties of the locally-emitted radiation 
do not change within the observable part of the stream. 
For example, all points within the beam may emit
a spectrum with the same shape but different normalisation. 
A case
 with weak spectral changes 
in the $\theta$ direction
is shown in Fig.~\ref{mili}.
2) When the streams/beams are extremely narrow
and emit the fixed-intensity patterns in the form of narrow ellipses such as
shown in Fig.~\ref{specgrad}c and \ref{mili}.
Then the arbitrarily-large angular separations of the fixed-$\nu$ contours
cannot much change the position of
a component. 
If the spread of contours in $\theta$ is very large, 
the component will just disappear. 
3) When a single-component
 profile is produced by a meridional stream ($\dcut \sim
90^\circ$); in this case the component's peak position
is $\nu$-independent, although the component's width may vary with frequency.

We conclude that the merging of components with increasing
frequency observed in T, Q and M pulsars, as well as the
frequency evolution of their width, can be naturally interpreted within the
fan beam model. The smaller rate of `RFM' observed
for the inner pair of components in Q and M pulsars, results from
the larger cut angle $\dcut$. It is not related to the weaker flaring
of the hypothetical emission region located at 
the inner polar magnetic flux tube.

The multistream symmetry of Fig.~\ref{rfm} is consistent with the ubiquitous
(though approximate) left-right symmetry of profiles: the outer components
are (on average) cut at a larger $\theta$ from the dipole axis.
If all streams are spatially
equidistant from the dipole axis (ie.~have the same footprint parameter $s$) 
then the inner components are
viewed at smaller $r$. They can therefore differ from the outermost 
components, having different flux, spectrum, or shape. The same argument
refers to the central component (`core'), which is a stream
viewed closest to the dipole axis. The stream model then retains the
capability of conal model in reproducing the roughly mirror-symmetric 
properties of profiles.

Note, that the mechanism of the $\nu$-dependent displacement of 
components, illustrated in Fig.~\ref{rfm}, works whenever
a spectrally non-uniform beam is cut at $\dcut \ne 90^\circ$,
regardless of whether the spatial RFM exists or not.
However, the width of the majority of profiles increases with $\nu$ (Hankins
\& Rankin 2010), which seems natural for the upward decrease of plasma
density.
The spatial RFM also provides
a way for symmetry breaking by the AR, as described in Sec.~\ref{arcentro}.
Therefore, it is reasonable to relate the $\theta$ spectral gradient
with the changes of the emitted spectrum along the stream, ie.~with
different emission altitudes.

The stream model is then capable to produce the apparent RFM effect.
A hint of this capability is visible in Figs.~2 and 3 
of DR12,
where the peak of a bifurcated component jumps to a later pulse
phase with increasing frequency. In the case of a single-peaked
(non-bifurcated) component, its
skewness would change and its peak would move rightward.
The apparent RFM can be especially strong when
streams are not too thin and when they are cut at small $\dcut$.

The choice of observed frequency $\nu=\nu_0$ picks up specific contours in 
Fig~\ref{rfm}.
These $\nu_0$-circles in different streams can be 
located above or below the horizontal
sightline path in Fig.~\ref{rfm}. 
If they are located above the path in the outermost streams,
then the peak flux of outermost components will increase
with decreasing $\nu$.
If at the same time the corresponding $\nu_0$-contours
in the inner streams (including a possible `core stream'), 
are located below the sightline path, a decrease of observed $\nu$
will make the peak flux of the inner components smaller.
Therefore, a change of observation frequency may reveal
different spectral evolution of the inner pairs (or core) 
than that of the outer pair (eg.~a brightening of the inner components
at a simultaneous fading of the outer pair).

Obviously, for a detectable RFM to appear, 
the streams have to be spectrally-nonuniform.
In general, different streams may have different $\theta$ spectral gradient.
In Fig.~\ref{specgrad} this effect is illustrated in the form of 
more overlapping and
more elongated fixed-$\nu$ contours.
This would result in different rate of component dislocation (and fading) 
with frequency, even when they have the same $\dcut$ (as is the case in 
Fig.~\ref{specgrad}).
Therefore, to obtain a more complete view of pulsar emission, it is necessary
to allow for the convolution of the spectral gradient with the effects of
$\dcut$.
Differences in the spectral gradient likely exist in pulsars and contribute
to the large variety of the observed behaviour of profiles
(Hankins \& Rankin 2010; Kramer et al.~1998).

One can imagine a model,
in which the spectral gradient changes monotonously 
with the azimuthal distance $\phm$ from
the MM. When convolved with the effect of $\dcut$ shown in
Fig.~\ref{rfm}, this could change the rate of, stop, 
or reverse the apparent RFM.
While such azimuthal spectral non-uniformity 
is possible, it does not automatically implies that the 
outer pair of components
should exhibit stronger apparent RFM. The value of $\dcut$ then remains
the primary and unavoidable factor, which determines the relative rate 
of components' displacement. However, if the spectral differences 
between the beams are strong, they need to be superposed 
on the effect of $\dcut$.

\subsection{Weakness of the apparent radius-to- frequency mapping 
in millisecond pulsars}

\begin{figure}
	\centering
	\includegraphics[width=0.47\textwidth,]{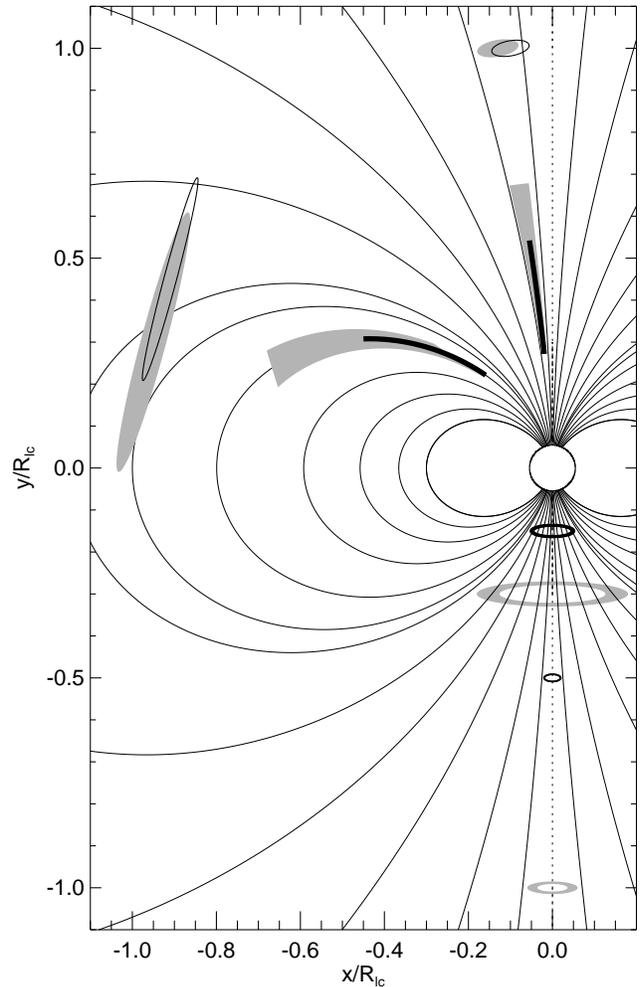}
\caption{Approximate geometry of emission regions at two different
frequencies. Bottom half of the plot presents the conal model whereas 
the top half -- the stream model. Lower $\nu$ is marked 
in grey, higher $\nu$ -- in black. Only the strongest-emissivity 
regions are marked for the conal model. 
Note the strong apparent RFM implied by the conal model for MSPs.
For the stream model, the
sky-projected beams are schematically presented on the far left
(MSPs) and top (normal pulsars). See Figs.~\ref{specgrad}c and
\ref{specgrad}b, respectively, 
for a more clear presentation of the beams.
}
\label{pers}
\end{figure}

The key difference between the millisecond pulsars (MSPs) and normal pulsars
is that B-field lines flare (bend away from the dipole axis) 
more strongly in the former case. 
For a fixed scale of radial spectral gradient, 
assumed to be universal for
both the MSPs and normal pulsars, the traditional conal model
would imply much stronger `RFM' for the MSPs.  
However, if the radio frequency corresponds to the plasma frequency,
emission at two frequencies should occur at 
radial distances which are at a fixed ratio,
as determined by the frequencies. 
In that case an apparent RFM
of comparable magnitude would be expected for both the MSPs 
and normal pulsars.
In the bottom half of Fig.~\ref{pers}, 
conal emission regions for both types of pulsars
have been illustrated for two frequencies at the same ratio of $r$. 
Available observational data (eg.~Kramer et
al.~1999; You et al.~2007) clearly contradict this expectation 
and show that MSPs  
do not exhibit the apparent RFM.

However, the lack (or weakness) of the `RFM' in MSPs is naturally expected
in the stream model. 
With more bent B-field lines, the streams in
magnetospheres of MSPs are likely to produce beams that are 
more elongated in the $\theta$-direction, 
and are more likely to be cross-cut by the line of sight.
More importantly, however, 
 changes of the spectrum with altitude
are spread over a very elongated beam. 
Therefore, 
the spectrum emitted at different points in the vicinity of the line of sight, 
does not change much with $\theta$. 
This corresponds to the overlapping and elongated fixed-$\nu$ patterns
shown in  Fig.~\ref{specgrad}c.
In the case of normal pulsars, the radiation from 
within a much larger range of altitudes, ie.~with a larger
spectral diversity, is projected onto a much smaller
angular area on the sky (see the top part of Fig.~\ref{pers}).
Therefore, the spectrum changes on a smaller angular 
scale in the beam of normal pulsars.
This corresponds to the less overlapping fixed-$\nu$ contours such as those
in Fig.~\ref{specgrad}b.
 
In the case of the MSPs, radiation emitted in a specific direction
has a larger spectral extent (is more broadband) 
than in the normal pulsars.
 While a stream of a MSP is being cut by the line of sight, 
the observer detects radiation from only
a small portion of the stream's extent, because the radiation is emitted
within a small angle $1/\gamma$ around the velocity vector ($\gamma$
is the Lorentz factor) and the curvature radius $\rho$ of the B-field lines 
is small. 
In the case of normal pulsars, with small curvature of B-field lines,
at any instant the observer probes a larger
range of $r$.  
However, this larger $\Delta r$ corresponds to a smaller range 
of plasma density, which can be shown as follows.
In a dipolar field, the range of plasma density $\kappa_{pl}$ is related to
$\Delta r$ through $\Delta\kappa_{pl}\propto r^{-4}\Delta r$, where $\Delta
r\propto\rho/\gamma$, as implied by the size of the microbeam. 
For the curvature radiation with the peak frequency
$\nu \propto \gamma^3/\rho$, the microbeam size is $1/\gamma\propto
\rho^{-1/3}$.
In a dipolar field $\rho \propto (r\rlc)^{1/2}\propto
(rP)^{1/2}$, so $\Delta\kappa_{pl}\propto r^{-11/3}P^{1/3}$.
In MSPs $r$ is two orders of magnitude larger, and $P$ -- two orders
of magnitude smaller, than in normal pulsars.
The range of plasma densities implied by the microbeam-size limitation
is then much larger in the case of MSPs than in normal pulsars.
The spectrum emitted in some direction $(\phm, \theta)$ is, therefore, wider
in the case of MSPs.

Another constraint on the range of plasma densities probed at a fixed
$(\phm, \theta)$ 
is related to the instantaneous visibility of the emission
region in the case when the extent of the $1/\gamma$ microbeam is negligible
(emission strictly tangent to B-field lines). 
For simplicity, let us ignore the aberration-retardation 
effects (described in Sect.~\ref{ar}), 
the issue of obscuration (Sect.~\ref{obscur})
and the possible torsion of electron trajectories (hence of the stream shape). 
Then at a single pulse phase the line of sight can receive radiation
from within a single plane of fixed $\phm$ within the emiter.
The observer is simultaneously detecting radiation from all the points
within the plane, 
where the B-field is inclined at a single angle
$\theta$ with respect to $\mu$.
As shown in Fig.~\ref{visi}, in the dipolar field these points
are located along a straight radial line with a fixed magnetic colatitude
$\thm\approx(2/3)\theta$. Thick parts of the line, 
which overlap with the grey emission regions, are quasi-simultaneously
visible by the observer. Fig.~\ref{visi} shows two emission regions: the
bottom one for a fast millisecond pulsar ($P\approx5$ ms), and the upper one for
a bit slower rotator. It can be seen that for a similar thickness 
of the streams
in $\theta_m$ direction, the observer samples a smaller interval
of $r$ in the case of the MSP (bottom grey region).
 The above-discussed extent of the $1/\gamma$ 
microbeam affects this picture by providing some non-zero $\theta_m$-width
to the simultaneously detectable region. 
Despite the smaller $\Delta r(\phm,\theta)$, a similar estimate as 
for the microbeam size again implies more broadband emission 
in the case of MSPs.

\begin{figure}
	\centering
	\includegraphics[width=0.48\textwidth,]{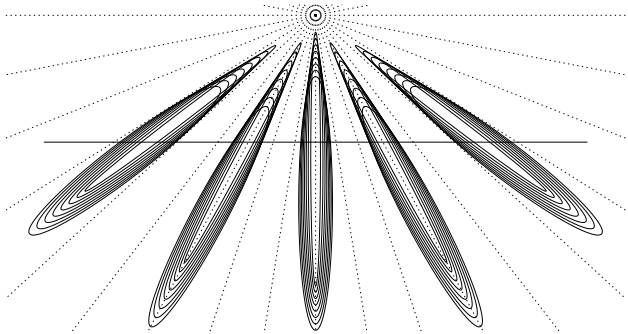}
\caption{Example of a millisecond pulsar beam. The fan beams are longer
and spectrally more uniform. Spectral changes in $\theta$ direction
are nearly washed-out as marked with the overlapping fixed-intensity 
contours for different $\nu$.  
The horizontal marks the passage of line of
sight. In this case pulse components do not 
spread with frequency.  
The AR effects, which bend the beams leftward (see Fig.~\ref{aber}b),
are neglected in this figure. 
}
\label{mili}
\end{figure}

\begin{figure}
	\centering
	\includegraphics[width=0.49\textwidth,]{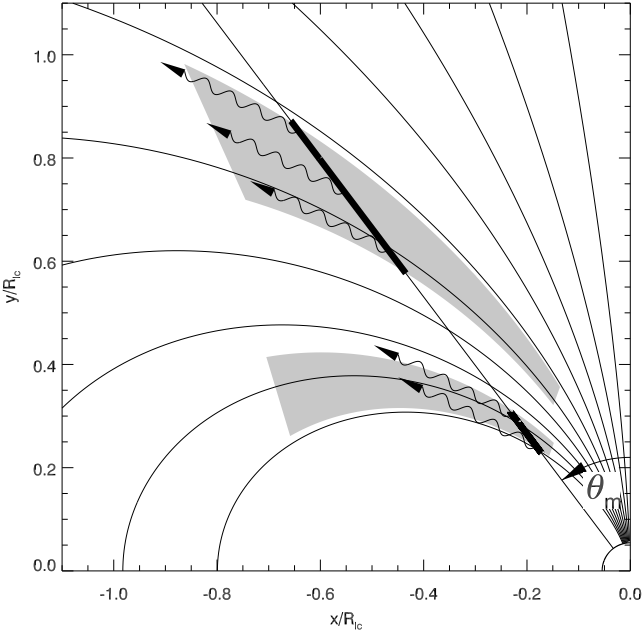}
\caption{Geometry of a region (tilted radial line) from which radiation is
detectable almost simultaneously. With emission constrained to the
grey contours,  the region becomes limited to the thick
sections of the radial visibility line. In the absence of AR effects, 
all the wavy rays should be observed at the same pulse longitude. 
}
\label{visi}
\end{figure}

For all these reasons (different length of detectable B-field line;
different angular extent of projected emission, and different
radial extent of detectable emission),  
the beams of MSPs are more elongated
and more broadband than those of normal pulsars. 
Cutting through the beams of millisecond pulsars then resembles
the situation presented in Fig.~\ref{mili}, with more elongated beams, 
and the spectrum weakly evolving along the beam. 
The wider frequency extent of the spatially more-localised spectrum
is consistent with the simpler shapes of MSPs' spectra, as described
in Kramer et al.~(1999), cf.~the spectra of normal pulsars in
Malofeev et al.~(1994), Kuzmin \& Losovsky (2001), and
Maron et al.~(2000).

\section{Mapping the apparent RFM back into the pulsar magnetosphere}
\label{mapping}

The problem of mapping the observed flux back into the dipolar 
magnetosphere
is a complex issue. The necessary mathematical formulae can be found eg.~in
DRD10. In general, however, even with a complete information
on the viewing geometry and location of the stream,
it would be difficult
to recover the $\nu$-dependent 
3D magnetospheric emissivity distribution
 from a pulse profile.
A useful partial insight can nevertheless be gathered from consideration of
possible meridional distributions of emissivity.

When the line of sight is crossing the stream,
it moves both in the magnetic azimuth $\phm$ and colatitude $\theta$
(Fig.~\ref{params}). 
The properties of polar zone (such as the acceleration potential) 
are expected to depend on magnetic azimuth $\phm$ and individual streams
may be asymmetric in $\phm$.
However, a localised plasma stream, ie.~a narrow stream
with large transverse gradient of plasma density, can possibly be considered
roughly symmetric with respect to its own central plane,
ie.~symmetric on both sides of the mid-stream azimuth $\phm=\phi_{m,0}$.
 Under this assumption, the plane of the fixed $\phi_{m,0}$
(containing the centre of the stream and the dipole axis)
becomes most usefull to apply the inverse engineering to `RFM'.
We will thus focus on the latitudinal motion of the sightline 
in $\theta$-direction because emission 
properties are expected to change monotonically with $r$ 
and $\theta$ (or $\thm$).
With the approximate $\phm$-symmetry assumed, 
the transverse motion of the line of sight
in the $\phm$ direction will just convolve a roughly symmetrical 
bell-shaped flux profile with the asymmetric flux changes resulting from
the motion in $\theta$.\footnote{Azimuthal density 
distribution may be more 
important for nearly meridional or very narrow streams,
however, since
they are expected to exhibit little `RFM', we will discuss 
the latitudinal density pattern. Some aspects of the lateral profile
in the $\phm$-direction are shown in Fig.~\ref{specgrad}. See also 
Sect.~\ref{obscur}.
}

\begin{figure*}
	\centering
	\includegraphics[width=0.8\textwidth,]{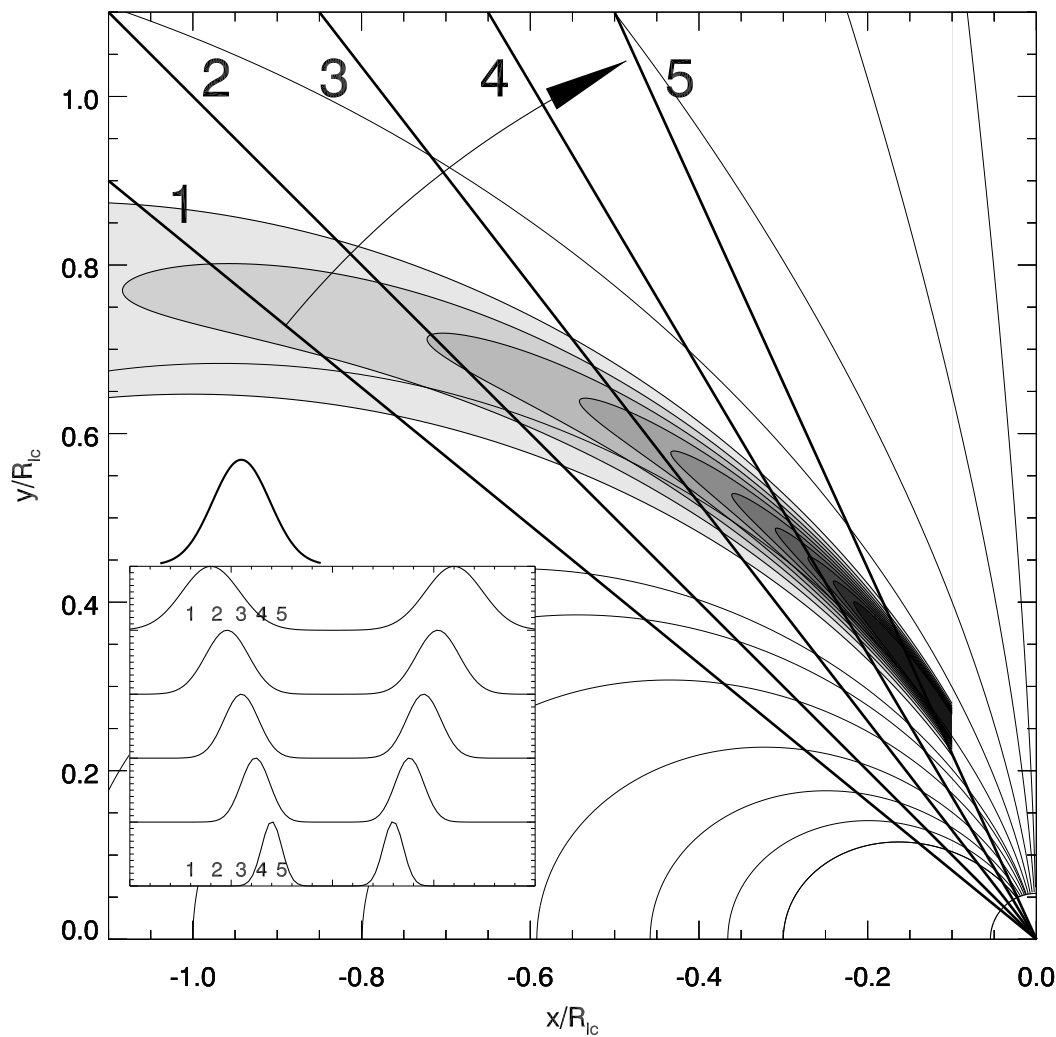}
\caption{Contours of equal density in the plane of the stream's central 
azimuth $\phi_{m,0}$. When the line of sight is passing through a leading-side 
stream,
it is probing increasingly deeper regions marked 
with the visibility lines 1-5, which
correspond to pulse longitudes marked in the inset. Therefore, 
the high-frequency emission dominates at later longitudes within 
the 1-5 pulse phase interval. The frequency 
increases from top to bottom in the inset. 
The bell-shaped curve above the inset presents the azimuthal
emissivity distribution within the stream, which needs 
to be convolved with the radial profile shown with the grey contours,
to estimate the locations of the components in the inset.
See text for more details.
}
\label{visi2}
\end{figure*}

Fig.~\ref{visi2} presents 
an emission region (stream) with a plasma density gradient
in the $\phi_{m,0}$-plane. 
Darker contours mark denser inner parts of the stream.
The contours of fixed density have been calculated for
a Gaussian lateral distribution and take into account
the 3D spatial divergence of B-field lines. Thus, the usual dipolar
decline of density with $r$ is included, though more sophisticated
complications (such as the null-charge surface) are neglected. 
It is assumed that the iso-density contours also
present the contours of fixed emissivity at a corresponding frequency,
with a lower frequency being emitted from a lower density region.
Finally, since we discuss averaged (or integrated) pulse profiles, 
the density distribution represents a long-timescale average 
of instantaneous density in pulsar magnetosphere.

Thick straight lines in Fig.~\ref{visi2} present five simultaneous-visibility 
regions
corresponding to five different pulse longitudes in the profiles shown in
the inset.
We assume that the stream extends at some angle $\phi_{m,0} < 0^\circ$
towards the leading side (LS) of the magnetosphere ie.~towards 
the direction of rotation (Fig.\ref{params}). 
The azimuthal extent of the stream ($\Delta\phm$)
is assumed to correspond to the pulse longitude interval 1-5.
In the absence of any radial (longitudinal) decline of emissivity,
the azimuthal emissivity profile which is perfectly symmetric with respect to
$\phi_{m,0}$,
would produce the almost symmetric
bell-shaped pulse component shown above the 
inset.\footnote{This azimuthally-shaped
component would not be perfectly symmetric, because the speed at which our
line of sight probes consecutive $\phm$ changes with pulse longitude, as
demonstrated by the S-swing of the polarisation position angle.
We ignore this effect to expose the role of the radial emissivity profile.}
To estimate the $\nu$-dependent behaviour of the component
in the case of the radially-nonuniform
emissivity, the radial emissivity profile needs to be convolved with
the azimuthal bell.

In general, our line of sight can enter 
the stream at
\emph{any} angular distance from the dipole axis, because this depends
on both the stream location and viewing geometry.
In the case shown in the inset, this happens
at such an angle $\theta$ that
corresponds to $\thm$ marked by the radial visibility line no.~1.
During the passage through the stream the sightline
approches the dipole axis and receives radiation 
from radial visibility lines with smaller $\thm$, ie.~located 
closer to $\vec \mu$. Near the pulse phase marked with the number `2' (see
the inset), a maximum of the lowest-$\nu$ component is observed, because 
the radial visibility line is detecting radiation from 
the bulk of the corresponding low-density region 
(the light-grey lobe that extends
to $x/\rlc\approx -1.07$).
%\footnote{Because 
%of the nonuniform plasma density, an 
%integration along the visibility lines for different pulse longitudes
% would be required to determine the actual pulse shape.
%However, this %would require specific assumptions about many
%% details of the emission region, which are 
%is not needed to present
%the main idea.} 
Note that only that part of the emission region which overlaps 
with the line no.~2, is detectable at the pulse longitude 
which is marked with number `2' in the inset.
When the line of sight is moving further through the pulse component
(to longitudes 3, 4, and 5),
the visibility line is progressing to locations 3, 4, and 5.
This increases the contribution of the high-$\nu$ radiation,
and diminishes the low-$\nu$ flux, as shown in the inset.
The line of sight can exit the beam
at any $\thm$ depending on the azimuthal width $\Delta\phm$ of the stream
and the observers' viewing angle. In the case illustrated 
in the inset of Fig.~\ref{visi2}, 
the sightline probes quite a large range of $\thm$ before exiting the stream. 
This happens when the beam is wide (large $\Delta\phm$)
or when it extends at a large azimuthal angle with respect 
to the meridional plane
($\phm \sim 90^\circ$, $\zeta\approx\alpha$).
Note that the flux decline in the wings of the discussed component
is caused by the azimuthal motion 
of the sightline.  
Even though Fig.~\ref{visi2} presents only the latitudinal motion, the effects
of the azimuthal motion on the shape of components are included in the
inset of Fig.~\ref{visi2}.

\begin{figure*}
	\centering
	\includegraphics[width=0.8\textwidth,]{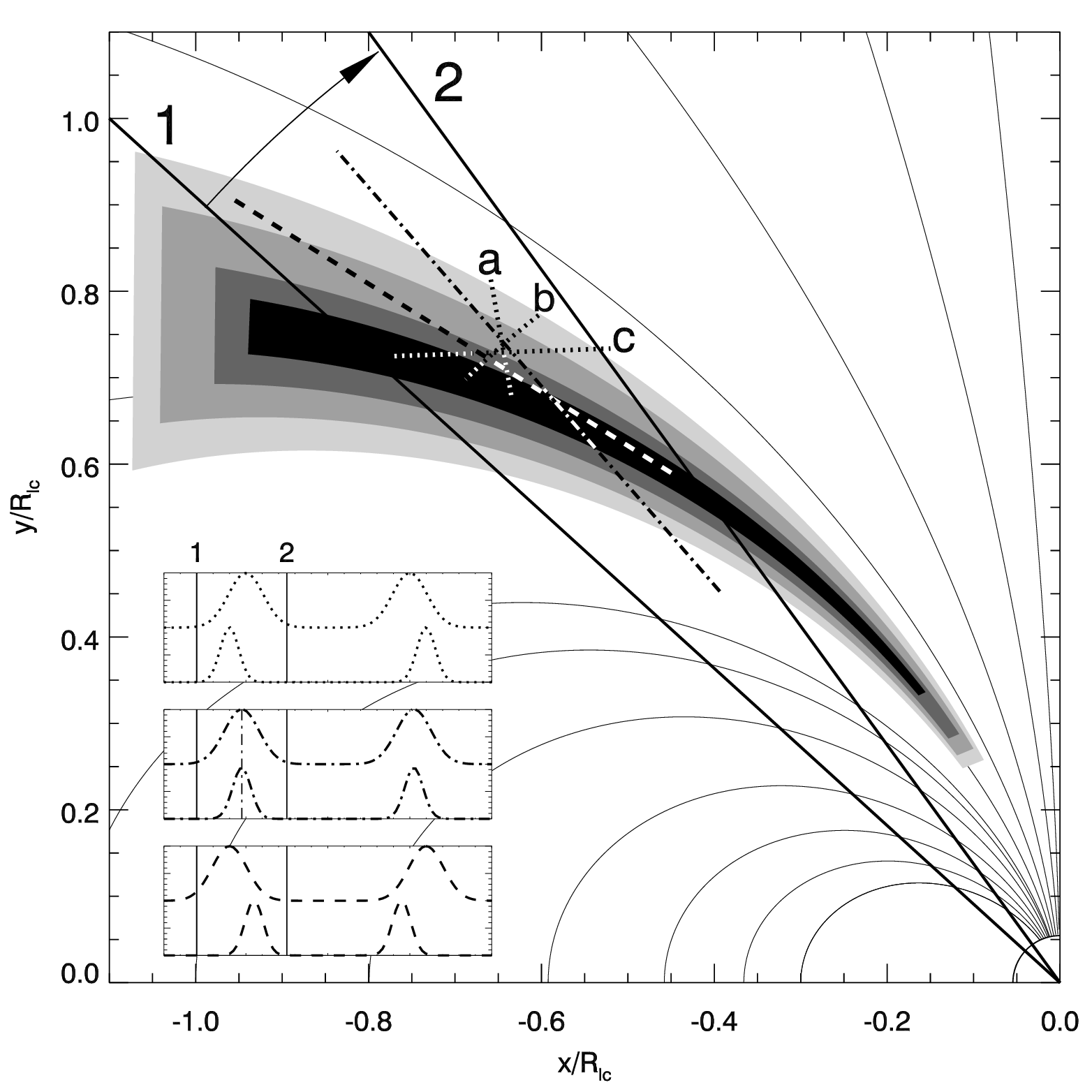}
\caption{The influence of spatial emissivity distribution on
the $\nu$-dependent displacement of components.
Grey contours of different grade show emission regions 
for different frequencies. Locations of peak emissivity are marked
with the dashed, dot-dashed, and dotted sections of straight lines.
The corresponding pulse profiles at two frequencies 
(with lower $\nu$ always shown on top) are presented in three insets.
The dot-dashed line separates the cases of normal `RFM'  from anti-`RFM'
behaviour.
See text for more details.
}
\label{visi3}
\end{figure*}

Since the stream we have considered was positioned on the LS,
our line of sight was \emph{approaching} the dipole axis during the
passage through the stream. 
To obtain the illusion of a cone, 
another component, located
 symmetrically on the trailing side (TS) of the profile would have to
be produced by another stream, located
roughly symmetrically on the TS of the MM.
Should such a stream exist, our sightline would
sample the visibility sections in a reversed order (5 to 1),
and produce the trailing component shown in the inset.

Depending on the geometry involved, 
at the moment when our line of sight is entering the stream,
the corresponding radial visibility line can have any angle 
with respect to the dipole axis.
The range of $\thm$ probed in Fig.~\ref{visi2} 
depends on the central azimuth of the stream $\phi_{m,0}$,
the width of the stream $\Delta\phm$, 
and the observer's impact angle $\beta$. For example, let us consider
a more meridional (or just narrower) stream with identical 
density distribution as before. The line of sight can enter the emission
region at the visibility line 3, and quickly exit the beam at the 
line no.~4. In such a case, the low-$\nu$ radiation would still be observed,
but the low-$\nu$ pulse component would become weaker, in comparison to the 
same-frequency components described before (as well as in comparison 
to the components at higher $\nu$). 
This is because the peak flux of the observed 
low-$\nu$ component must now be
located roughly mid-way between pulse longitudes `3' and `4', 
as determined by the new azimuthal emissivity profile
 (not shown in Fig.~\ref{visi2}).
 At this phase, 
however, the low-$\nu$ emissivity is weaker (see the top profile
in the inset of Fig.~\ref{visi2}). 
Therefore, the observed spectrum of a given component
is selected by the path of the line of sight 
within the stream
(in addition to being determined by intrinsic factors such as
the  emission physics). In the presented case, ie.~for emission 
altitude quickly increasing with decreasing $\nu$,
the outer components should have softer spectra
than the inner components,
ie.~the outer components should increasingly dominate 
in the profile at lower frequency. This phenomenon is observed
in several pulsars,  
eg.~in the case of 
B1237$+$25 (Hankins \& Rankin 2010).

A more general case is shown in Fig.~\ref{visi3}, in which
the transverse arrangement of emission regions that correspond to different
 frequencies (and densities) is shown as before 
with the grey-scale contours. However, it is now assumed that the local
emissivity within each fixed-$\nu$ region changes with altitude. 
Five possible radial (more precisely -- longitudinal) profiles of emissivity
are considered within the stream. 
In each case
the local emissivity
increases along the stream until reaching a maximum at some $r$, 
then falls off with further increase of the distance along the stream.
For each case,
the maximum of emissivity within the stream is  marked
with
sections of straight lines: dashed, dot-dashed, and dotted (three cases).
We will discuss them in the clockwise order, starting from the dashed case.
Pulse profiles that correspond to different cases are presented
in the three insets in the bottom left corner. 
In each of them the low-frequency profile is on top.

The sightline
is assumed to enter the stream from a side at pulse phase `1',
corresponding to the visibility line no.~`1'.
Then the sightline moves obliquely through the stream and exits it at 
phase `2', associated with the visibility line no.~'2'. 
The low-density region that overlaps with a visibility line is split in two
parts, one on the top, another on the bottom side of the stream. The
question on whether the radiation from the bottom part can reach the
observer is difficult and depends on several factors, such as the
polarisation of the emitted wave, and the temporal/spatial variations of
plasma density in the stream. A qualitative inclusion of the bottom-side
emission would require an interesting, but lengthy analysis of a more
complicated version of Fig.~8. For brevity, therefore, 
it is now assumed 
that the bottom side of the stream 
(the part of the stream on its concave side) contributes little to the
observed flux. 

The dashed line presents the case when the altitude of peak emissivity
steeply decreases with increasing frequency.
A cut through a stream with such a property would result in normal `RFM',
shown in the bottom inset of Fig.~\ref{visi3}. This is because
the strongest low-$\nu$ emission is observed early in the 1-2 
phase interval, when the observer is sampling visibility
lines located close to the line no.~`1'. High-$\nu$ emission
will peak closer to phase `2', since the high-$\nu$ emissivity 
is strongest nearby the visibility line no.~'2'.
We remind that the decline of flux at the edges of the 1-2 interval
is caused by the azimuthal emissivity profile, which needs to be convolved
with the radial profile peaking at the dashed line. 

\begin{figure*}
	\centering
	\includegraphics[width=0.8\textwidth,]{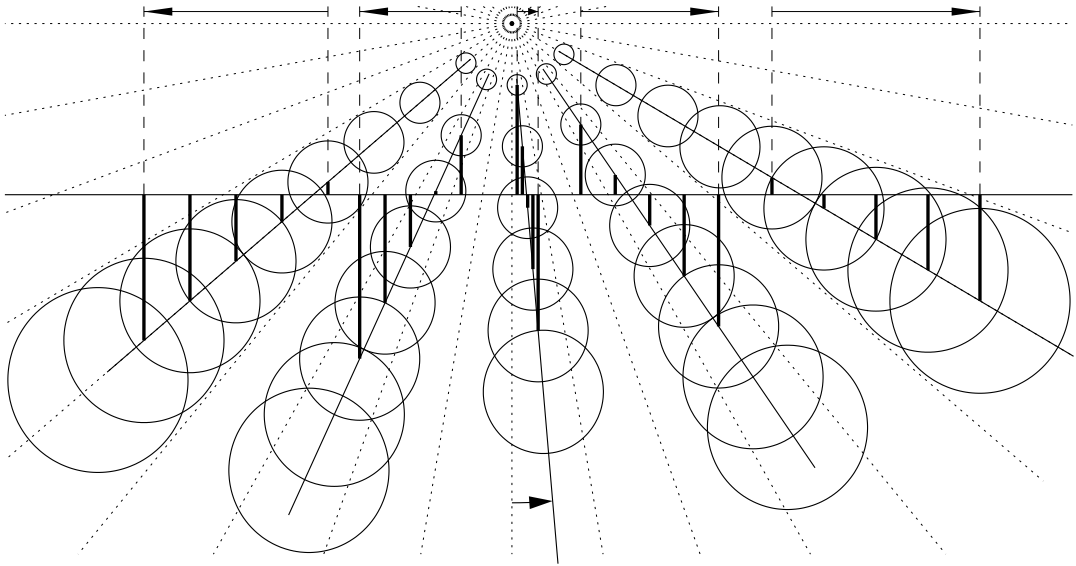}
\caption{The mechanism of the accidental lag of the 
central component with respect to
the centroids of pairs of outer components. 
The beam structure is symmetric with respect to the central beam, 
but it is rotated by the angle $\epsilon$ (bottom arrow)
%, $\epsilon$ is assumed to be positive) 
with respect to the MM.
All the beams have the same $\theta$ spectral gradient.
The arrows on top show the components' displacements for decreasing $\nu$.
The model produces a lag of the central component 
which is increasing with decreasing $\nu$ (see text).
}
\label{centroid}
\end{figure*}

The dot-dashed line presents the fixed-$\thm$ case when all frequencies
peak roughly in the middle of the 1-2 phase interval. This would not
induce much `RFM', because the convolution of two coincident bell-shaped
profiles does not change the peak's position (see the dot-dashed
components in the middle panel of the inset). Thus the dot-dashed section
is a separatrix between the normal `RFM' and 
the inverse `RFM', in which case the lower-$\nu$ profiles would be observed
closer to the centre of the profile.\footnote{The lack of `RFM' 
is a frequently 
observed feature of several pulsars (eg.~Gould and Lyne 1998).
It can be interpreted through
the coincidence of the peak-emissivity separatrix with the centre of an
observed component. Another explanation
 involves a quasi-uniform longitudinal emissivity distribution 
within the entire 
region between the visibility lines `1' and `2'.}
The speed at which the line of sight crosses consecutive azimuths $\phm$
increases towards the MM, as has always been 
illustrated by the S-swing of the polarisation position angle. 
Therefore, the separatrix that corresponds precisely to the centre of the 
1-2 pulse interval is located not exactly in the middle between
the visibility lines `1' and '2'. For the leading component 
shown in Fig.~\ref{visi3}, the separatrix is closer
to the visibility line `2'.
 
An interesting case leading to the inverse  apparent RFM is represented 
by the first dotted line, marked `a'. The spatial RFM is normal in this case, 
with the peak emissivity located at a larger $r$ for the lower-$\nu$ 
region. However, the low-$\nu$ radiation peaks at a later
phase (closer to the visibility line `2') than the high-$\nu$ emission,
that peaks closer to phase `1'. This is then the case when the normal spatial
RFM leads to the inverse apparent RFM, illustrated with the dotted profiles
in the top inset.

The second dotted line, marked `b', presents the case with no
spatial RFM. The peak emissivity at all 
frequencies occurs at the same radial distance $r$.
The line `b' is then
orthogonal to the separatrix described above.
Orientation of the dotted line with respect to the visibility lines `1'
and `2' reveals that
the low-frequency component (top curve in the top inset)
will definitely peak closer to phase `2', whereas the high-$\nu$ component
closer to phase `1'. This is the case when the lack of spatial
RFM results in the inverse apparent RFM.

The nearly horizontal dotted line `c' presents the case
with the altitude of peak emissivity steeply decreasing with
decreasing $\nu$. It produces a similar profile evolution 
as in the previous case (top inset).

Thus, the frequency-dependent  behaviour of profiles (`RFM', no `RFM', or
inverse `RFM'), depends on the spatial 
emissivity distribution around the fixed-$\thm$ separatrix 
that corresponds to the central azimuth of the stream $\phi_{m,0}$.
Since the majority of pulsars exhibits normal RFM, it is concluded that
typically the altitude of the strongest emissivity at a given frequency
quickly increases with decreasing $\nu$, 
as the straight dashed line shows qualitatively.
This is consistent with the common expectation of plasma density drop with
altitude illustrated by the contours in Fig.~\ref{visi2}. 
Interestingly, however, it is possible to observe the reverse RFM
even for the peak-emissivity altitude increasing with decreasing $\nu$
(case `a').
%This would occur for all cases with a peak emissivity line
%more vertical than the dot-dashed (fixed-$\thm$) separatrix and 
%the short dotted (fixed-$r$) line
%in Fig.~\ref{visi3}.

\section{Origin of the fre\-quen\-cy- -de\-pen\-dent
lag of the central component}
\label{ar}

Position of a central component in multiple-component profiles
often lags a centroid located half way between outer components.
%(hereafter this central point of a pair will be called a `centroid').
GG01 have shown that this can be caused by the 
aberration and retardation (AR) effects 
operating at different emission altitudes of the core, inner cone, 
and outer cone.
The AR-induced shift of a centroid 
is approximately equal to $\Delta\phi\approx
2\Delta r/\rlc$, ie.~it depends only on differences in emission altitude
(Dyks et al.~2004).
It has therefore been widely used to determine $r$ 
as well as the transverse distance of the cones from the dipole 
axis, ie.~the footprint parameter $s$
(GG01; GG03). 
However, the ensuing structure of the radio emission region (with
a variety of $r$ and $s$) have revealed unacceptable arbitrariness, 
difficult to interpret in terms of physics.  
For example, instead of the lateral separation of emission rings
(Ruderman \& Sutherland 1975; Wright 2003),
for PSR B0329$+$54
GG01 invoke as many as four radially-separated
 emission cones at roughly the same B-field lines ($s\sim0.6$).
As found in GG03 (and adjusted in Table 1 of Dyks et al.~2004),
the set of the hypothetical emission rings can be located 
near the dipole axis 
($s\in(0.3,0.4)$, PSR B2111$+$46), distributed around a moderate
distance from the pole ($s\in(0.3,0.6)$, B1237$+$25) 
or placed close to the polar cap rim
($s\in(0.7,0.9)$, B1857$-$26).

As soon as one allows for the azimuthal structure of the radio emission
region, the centroid shift can be understood without invoking 
emitters at disparate altitudes.
As we show below, with several streams extending within
 the same altitude range, the shift can still be understood in two ways:
1) through the altitude-dependent AR effects that break
the symmetry of the system of streams (Sect.~\ref{arcentro}), and
2) just through an accidental asymmetry of location of the beam system
with respect to the MM, whatever
is the reason for the asymmetry. In the latter case,
discussed in the next subsection, the centroids are shifted
regardless of emission altitudes, ie.~the shift is non-zero
even for exactly the same emission altitude within the entire emission beam
($r=r_0$, $\Delta r =0$). 
The AR effects are not involved in this mechanism.
The key factor that leads to the centroid
shift in both mechanisms, is the asymmetry of streams/beams
with respect to the MM, with the
cut angle $\dcut$ again responsible for the frequency-dependence
of the phenomenon.

 Below we test the two mechanisms against the observed properties of the
centroid shift phenomenon. We start with the simple case with no AR
effects included.

\subsection{Centroid shift
due to the asymmetric location of the beam system}
\label{accidental}

\begin{figure}
	\centering
	\includegraphics[width=0.48\textwidth,]{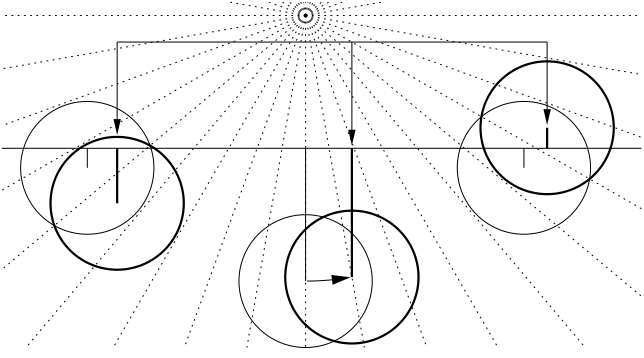}
\caption{A cartoon showing that the backward misalignment
(counterclockwise rotation around $\vec \mu$) 
of a symmetric system of circular beams
results in the lag of the central component
with respect to the centroid of the flanking pair.
Thick lines present the misaligned case. The horizontal line marks the path of
the sightline.
}
\label{misal}
\end{figure}

Since the pulsars considered by GG03
exhibit approximate symmetry of their profiles,
we consider a beam system which is perfectly symmetric
with respect to the central plane
($\phi_{m,0}$-plane) of its central stream.
To emulate the asymmetry, the beam system is rotated by a small angle
$\epsilon$ with respect to the MM.
% ($\epsilon$ is assumed to be positive).
% (Fig.~\ref{centroid}). 
Such a system with five identical beams is presented in 
Fig.~\ref{centroid}.
Consecutive circles along each beam again present 
the fixed-intensity
contours at different and equally separated $\nu$.
All the contours are also equally separated in $\theta$, 
ie.~the $\theta$ spectral gradient is assumed to be the same
for all beams. 
The misalignment of the system with respect to the MM produces
a lag of the central component with respect to the centroids
of the flanking pairs.  
This is because in the considered case
 (flat geometry with $\alpha \sim 90^\circ$)
the observed pulse longitude of a component is given by:
\begin{equation}
\phi = \theta_\nu \sin\phm
\label{faza}
\end{equation}
where the angle $\theta_\nu \equiv \theta(\nu)$ corresponds 
to the fixed-intensity contour
at the observed frequency $\nu$, (hereafter the index $\nu$ will be omitted),
and $\phm$ is the azimuth of the stream (note that $\phm$ is negative
on the LS and positive on the TS).
%where $\beta$ is the impact angle (Fig.~\ref{params}).
If the same-$\nu$ contours in different beams are located at the
same $\theta$, the
slower-than-linear increase of $\sin\phm$ ensures that the LS
spaces between adjacent components are larger 
than the `corresponding' TS spaces (Fig.~\ref{misal}).

A simple mathematical model, presented in the next subsection
(\ref{mathacc}) reveals that the properties of the 
model shown in  Fig.~\ref{centroid} are consistent with the
observed $\nu$-dependent characteristics of the centroid shift. 
The model can simultaneously explain the normal apparent RFM,
the sign (direction) of the shift, the increase
of the shift with decreasing $\nu$, and the increase of the shift 
towards the profile periphery.

However, the misalignment of the beam system 
has no built-in preference for the \emph{leftward} shift of the centroids.
A mirror image of Fig.~\ref{centroid} would produce exactly the opposite
effect -- the lag of the centroids with respect to the central component.
Moreover, the effect must be affected by the AR shift, 
as soon as there are
altitude differences associated with detectable parts of different streams
(see Sect.~\ref{arcentro}).

\subsubsection{Simple model for the accidental shift of centroids}
\label{mathacc}

The mathematical model presented in this section
assumes the circular fixed-$\nu$ patterns shown in
Fig.~\ref{centroid}, ie.~the case shown in Fig.~\ref{specgrad}b
is modelled with the case shown in Fig.~\ref{specgrad}a. 
We can define an intra-system magnetic azimuth $\Phi$ 
measured around $\vec \mu$ 
from the central beam of Fig.~\ref{centroid}, so that $\phm = \Phi+\epsilon$.
The beam system is symmetric in $\Phi$. 
The longitudes of the LS and TS
components then become: $\phi_{\sss L}=\theta\sin(-|\Phi_{\sss L}|+\epsilon)$
and $\phi_{\sss T}=\theta\sin(\Phi_{\sss T}+\epsilon)$, respectively. 
The LS azimuth is assumed to be negative, so we use
the absolute value to expose the magnitude of $\phi$.
For our perfectly symmetric beam system $|\Phi_{\sss L}|=\Phi_{\sss T}=\Phi$.  
The location of a pair centroid is then 
\begin{equation}
\phi_{\rm pair} = \frac{\phi_{\sss L}+\phi_{\sss T}}{2} =
\theta\cos\Phi\sin\epsilon.
\label{centro}
\end{equation}
The central component (`core') is located at the pulse phase
\begin{equation}
\phi_{\rm ctc}=\theta\sin\epsilon.
\label{coreposnoar}
\end{equation}
Assuming the same $\theta(\nu)$ for both the central and flanking
components, the centroid precedes the `core' by the pulse phase interval
\begin{equation}
%\phi_{\rm pair} - \phi_{\rm ctc} = -\beta\tan\epsilon \ 
%\frac{\tan^2\Phi(\tan^2\epsilon + 1)}{1 - \tan^2\Phi\tan^2\epsilon}.
\Delta\phi_{\rm obs} = \phi_{\rm pair} - \phi_{\rm ctc} = 
\theta\sin\epsilon(\cos\Phi - 1) \ < \ 0.
%\frac{-\beta\tan\epsilon\tan^2\Phi(\tan^2\epsilon + 1)}
%{1 - \tan^2\Phi\tan^2\epsilon},
\label{lagnoar}
\end{equation}
For pairs of more peripheric components at a larger $\Phi$
(smaller $\cos\Phi$) the magnitude of $|\Delta\phi_{\rm obs}|$ increases
and so the lag of the central component. This is consistent
with the properties of pulsars studied by Gupta and Gangadhara.
The lag increases with the misalignment $\epsilon$ of the beam system
(eq.~\ref{lagnoar}).

Thus, for a bunch of similar beams with similar angular spectral gradient
in the $\theta$-direction, the misalignment alone
can produce the forward shift of centroids, 
which is larger for the outer pairs of components. 
However, GG03 also show that 
in the observed profiles the flanking pairs 
of components precede the `core' by a phase interval which increases
with decreasing $\nu$. 
To test if the `accidental asymmetry' model of Fig.~\ref{centroid} 
can explain this,  
we will 
use the following simple mathematical model.
%the behaviour of the elongated patterns 
%(Fig.~\ref{specgrad}b) will be modelled with 
%the
%simple circular case of Figs.~\ref{centroid} and \ref{specgrad}a.  
 
%In the case of the circular patterns, 
%%(cf.~different panels in  Fig.~\ref{specgrad}),
%% Fig.~\ref{centroid}  
%and assuming
With
 the flat and equatorial approximation 
(small beam, $\alpha\approx90^\circ$),
%At the fixed $\phm$, the spectral properties are the function of $\theta$,
%In the case of the circular fixed-$\nu$ contours (cf.~different panels in 
%Fig.~\ref{specgrad}),
%Considering only the spectral properties along $\theta$
The observed pulse longitude of a peak of a component 
is moving with the frequency at a rate: %$d\phi/d\theta$:
\begin{equation}
\frac{d\phi}{d\nu} = \frac{d\phi}{d\theta}\frac{d\theta}{d\nu} =
\frac{d\phi}{d\theta} \dsp =  \dsp \sin(\phm)
\label{dede}
\end{equation}
where $\dsp=d\theta/d\nu$  
sets
%the angular spectral `gradient' 
the angular scale of spectral variation 
which corresponds to the
displacement of the fixed-$\nu$ contours in the $\theta$ 
direction.\footnote{Note that if $\dsp$ is considered as a free parameter, 
then the mathematical model is valid for any fixed-$\nu$ patterns, for which
$d\phi/d\nu \propto d\phi/d\theta$.
}
In general,
$\dsp$ may change along the beam and can 
be different for beams located at different $\phm$. 
%In our case $\dsp$ is the same for all beams and independent of
%$\theta$, therefore, hereafter it will be normalised to the value of $-1$
%(where the minus sign reflects the `normal RFM', ie.~the 
%increasing distance of the lower-$\nu$ patterns with $\theta$, whereas 
%the plus sign is for the anti-RFM case).
 
Below we consider the negative of derivative \mref{dede}
to learn how the profiles
evolve with decreasing frequency ($-d\nu=\nu_{\rm low}-\nu_{\rm high}$). 
Accordingly, with \emph{decreasing} frequency, 
the central component moves rightward 
at the rate:
\begin{equation}
%\left(\frac{d\phi}{d\theta}\right)_{\rm ctc} = 
v_{\rm ctc} = \frac{d\phi_{\rm ctc}}{-d\nu}%\right)%_{\rm ctc} 
= -\frac{d\phi_{\rm ctc}}{d\theta} \dsp = 
\dspc \sin{\epsilon} \approx \dspc \epsilon
\label{corenoar}
\end{equation}
where the index `ctc' stands for the `central component',  
and we have substituted $\dsp=-\dspc$ for the $\theta$ spectral scale
of the central component.
% (in the vicinity the sightline-crossing point). 
The minus sign is
 supposed to reflect the
normal RFM, for which $d\theta/d\nu < 0$ and $\dspc>0$. 
For a pair of flanking components (inner or outer) 
we have $d\phi/d\theta=
\sin(\Phi_{i} + \epsilon)$, where the index `$i$' refers 
either to the leading ($i=L$) or trailing
($i=T$) location with respect to the central beam.
Therefore, the speeds at
which the components on the LS and TS departure from the
MM at \emph{decreasing} frequency are:
\begin{equation}
v_{\sss L} = %\left(
\frac{d\phi_{\sss L}}{-d\nu} %\right)%_{\sss L} 
= \dspl \sin(-|\Phi_{\sss L}|+\epsilon) 
\end{equation}
\begin{equation}
v_{\sss T} = %\left(
\frac{d\phi_{\sss T}}{-d\nu}%\right)%_{\sss T} 
=  \dspt \sin(\Phi{\sss T}+\epsilon),
\end{equation}  
where $\dspl$ and $\dspt$ set angular the spectral scales
for the corresponding 
flanking components, ie.~$d\theta_i/d\nu = -\Delta^{\negthinspace i}$. 
For the symmetric properties of the beam system ($|\Phi_{\sss L}|=
\Phi_{\sss T}=\Phi$, $\dspl=\dspt=\dspp$, Fig.~\ref{centroid}), 
the pair's centroid moves rightward at the speed
\begin{equation}
v_{\rm pair} = \frac{v_{\sss L} + v_{\sss T}}{2} = 
\dspp \sin\epsilon\cos\Phi \approx \dspp \epsilon\cos\Phi.
\label{vel}
%\ > \ \left|\frac{d\phi}{d\theta}\right|_{\rm ctc} 
\end{equation}  
If the spectral properties 
of the central and flanking subbeams are the same
($\dspc = \dspp$), 
 the rightward speed of the centroid is 
smaller than the speed \mref{corenoar} of the central component, 
and the more so
for more peripheral pairs. 
Therefore, the `core' lag increases with decreasing 
observation frequency
(or, more precisely, with increasing $\theta$), in agreement with
the afore-mentioned observations. The rate at which the
centroid departures from the core is:
\begin{equation}
v_{\rm obs} = v_{\rm pair} - v_{\rm ctc} = 
-\sin\epsilon(\dspc - \dspp\cos\Phi).
\label{relvel}
%\ > \ \left|\frac{d\phi}{d\theta}\right|_{\rm ctc} 
\end{equation}
Since the rightward speed of a pair centroid (eq.~\ref{vel})
decreases towards the profile periphery, the centroids of more peripheral 
pairs  departure from the core at a larger rate.
This behaviour is consistent with observations.

Accordingly, 
the backward misalignment of the beam system, shown in Fig.~\ref{centroid},
can simultaneously explain all the $\nu$-dependent features 
identified for the pulsars of Gupta and Gangadhara:
1) the normal apparent RFM,
2) the lag of the central component, 3) the increase of the lag 
with decreasing $\nu$, and 4) the increase of the lag 
for more peripheric pairs.

In general, the $\theta$ spectral scale may depend on $\theta$ and
be different for different subbeams. As demonstrated by eq.~\mref{relvel},
the difference between 
$\dspc$ and $\dspp$ at the sightline crossing points can change the 
 magnitude and direction of the centroid shift. 
For example, for some physical reasons 
the $\theta$ spectral scale could increase with the azimuth $\phm$
in a slightly asymmetric way with respect to the MM.
This could produce a result identical to that in Fig.~\ref{centroid}  
even for a symmetric beam system centred at the MM.
However, such spectrally-based interpretations 
do not have an obvious shift-sign preference and
 fail to automatically justify the increase of the shift towards the
periphery of a profile (at a fixed $\nu$) as well as the increase of 
the shifts with  $\nu$.

The number of possible combinations of parameters
(different shape of the fixed-$\nu$ pattern, direction of the misalignment,
and the sign of spatial RFM) is larger than discussed above, 
and they result in different observable properties. For instance,
if the system of circular patterns is misaligned leftward with respect to
the MM (opposite than shown in Fig.~\ref{centroid}), 
then the pair centroid lags the central component by a phase interval
increasing with $\Phi$ (eq.~\ref{lagnoar} with a negative $\epsilon$).
For decreasing $\nu$, such a centroid would departure rightward 
from the `core' (for a negative $\epsilon$, eq.~\ref{relvel} 
predicts $v_{\rm obs}>0$).

For the elongated patterns of Figs.~\ref{mili} and \ref{specgrad}c (with
little or no apparent RFM), 
the location of components is determined by the points at which the sightline 
is crossing through the central azimuth of the beam
(eq.~\ref{fabta}). As we show in the
Appendix, 
%location determined by eq.~\ref{faza})
a system of such elongated beams would have to 
be misaligned leftwards (opposite than shown in
Fig.~\ref{centroid}) to produce the centroid advance. This type of a model
(with no apparent RFM)
would exhibit a frequency-independent lag of the central component.

The accidental asymmetry model shown in Fig.~\ref{centroid} 
is then consistent with the properties of pulsars described by GG03.
However, depending on the direction of the beam misalignment (the sign of
$\epsilon$),
it can produce both the leftward and rightward shifts. This may possibly
be inconsistent with the fact that pulsar literature
is dominated by descriptions of the leftward centroid shifts that increase 
with decreasing $\nu$ and with the proximity of profile periphery.
%(eg.~GG03). 
However, the rotation of a pulsar, 
through the AR effects, is able to induce both the
system's spatial asymmetry in $\dcut$  as well as some
$\phm$-asymmetry in the system's spectral gradient.
Therefore, we turn now
to the subject of how pulsar rotation
breaks the symmetry of the multiple-stream system
through the AR effects.

\begin{figure*}
	\centering
	\includegraphics[width=0.8\textwidth,angle=0]{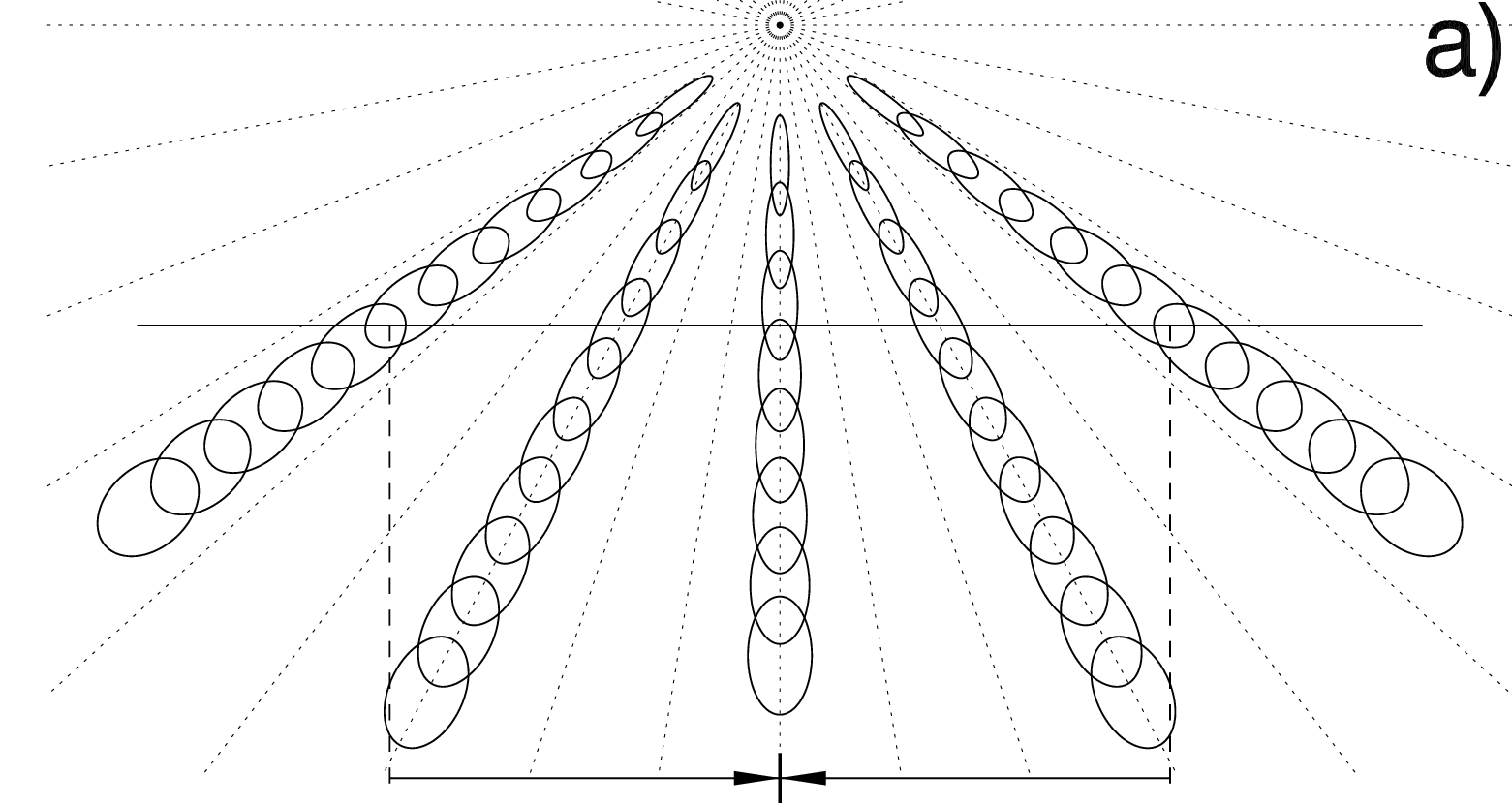}
	\includegraphics[width=0.8\textwidth,angle=0]{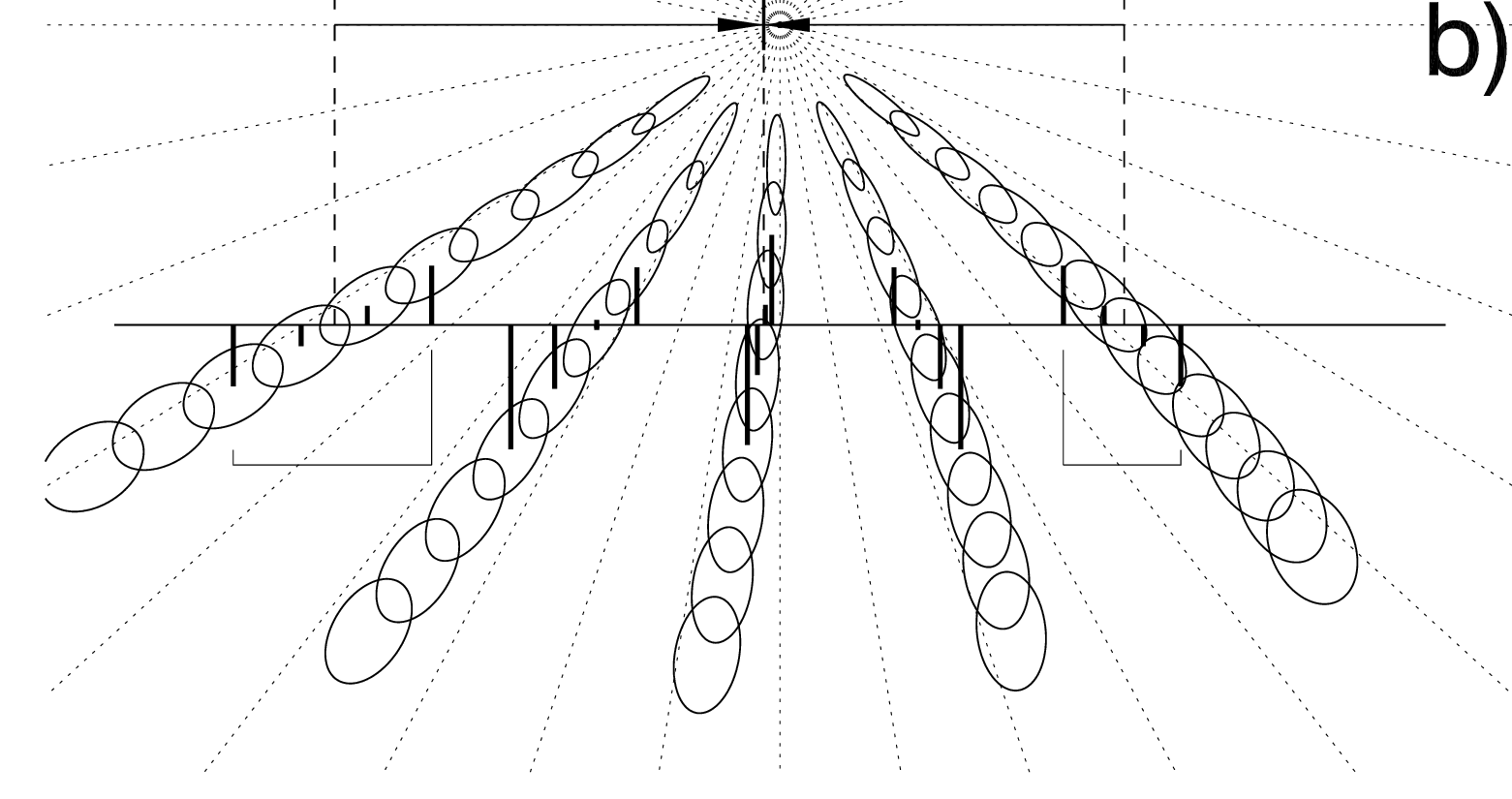}
\caption{The mechanism of central component's lag with respect to
centroids of outer components. {\bf a)} A sky-projected emission 
beam with no aberration-retardation effects included. 
The ellipses present fixed-intensity 
contours at different frequencies, which decrease outwards.
Emission altitude decreases towards the dipole axis.
{\bf b)} The beam with the AR-effects included. 
Note the forward bending of the individual beams.
All ellipses were shifted leftward by a $\theta$-dependent interval 
$|\Delta\phi|\propto r\propto\theta^2$.
The lag of the central component is marked with the horizontal arrows.
Note the decreased cut angle $\dcut$ and the increased 
width of the leading components. This smaller value of 
the leading-side-$\dcut$ 
implies a larger lag at lower $\nu$, 
 through the mechanism shown in Fig.~\ref{rfm}. The ellipses 
on the TS are
more overlapping with each other, which enhances the
effect of the centroid shift.
}
\label{aber}
\end{figure*}

\subsection{Shift of the centroids due to the AR effects}
\label{arcentro}

Fig.~\ref{aber}a presents the sky-projected beam 
that should be typical for the M pulsars. 
The perfect symmetry is assumed only to clearly expose
the phenomenon of the centroid shift. In general, we allow for a variety 
of asymmetries (in $\phm$, $\Delta\phm$, and $\theta$, see Fig.~18 in DRD10
and the illustrations of beams in Wang et al.~2014). 
However, the low-altitude dipolar 
magnetic field, as well as the electric
field within the polar tube, are expected to be
symmetric with respect to the MM.
Therefore, it is natural to assume that the left-right symmetry
should be on average found in pulsars.  
However, the AR-induced 
shift of centroids 
is expected to appear also for beam systems
which are misaligned with respect to the MM.

The elliptic contours in Fig.~\ref{aber}a
present the fixed-$\nu$ intensity pattern in the way similar to Fig.~\ref{rfm}.
The ellipses mark the uniform $\theta$ spectral gradient.
In reality they may
 be more elongated in the direction of the stream
(see Figs.~\ref{mili} and \ref{specgrad}bc),  
but this would make the figure unclear.
Emission at a fixed $\nu$ is not constrained 
to the interior of an ellipse only.  
The ellipses may present, let us say, 
 the $90$\% intensity level, 
with the  peak emissivity at the centre of each ellipse. 
The solid horizontal marks the passage of the line of sight, again
for a large dipole inclination and an overall  small angular extent 
of the beam (which allows us for the flat approxiation of a 
generally-spherical case).

The pulsars that exhibit the centroid shift also show 
the normal apparent RFM. The separation of their outer peaks increases
with decreasing frequency. Simultaneously, the centroid shift becomes
larger and the more so for the outer pair of components 
(GG01).
As has been shown in Sect.~\ref{mapping}, for pulsars exhibiting 
the normal apparent RFM, the altitude of emission region
must increase steeply with decreasing $\nu$.
The frequency associated with different ellipses in Fig.~\ref{aber} 
is then decreasing with the angular distance from the dipole axis.

Since circumpolar B-field lines are quasi-radial, 
all the fixed-$\nu$
ellipses undergo the leftward AR shift (in the direction of early
rotational phase) by $2r/\rlc$, irrespective of the dipole tilt $\alpha$ 
 (Dyks et al.~2004).
This is shown in Fig.~\ref{aber}b, in which the streams
are bent in the direction of rotation.
In a dipolar field the radial distance of emission $r$ increases with the
angular distance $\theta$ from the dipole axis according to $r\propto
\theta^2$
so the magnitude of the shift quickly increases along the beam.
Therefore, the central component is shifted leftward by the smallest 
interval of pulse longitude, because the sightline cuts the `core stream'
at the smallest $\theta$ and $r$. The inner pair of components
undergoes a larger shift, but not as large as the outer pair.
A result of a full 3D simulation of this effect 
is presented in Fig.~4 of DR12.\footnote{Note that the most 
vertical streams
in Fig.~4 of DR12 correspond to $\phm=170^\circ$ (the top one)
and $\phm=10^\circ$ (the bottom one), ie.~the `core beam'
(corresponding to the stream within the MM) has been replaced
with a second neighbouring beam of a stream located on the TS.}

As can be seen in Fig.~\ref{aber}b, the AR effects break
the beam symmetry in such a way that the cut angle $\dcut$ decreases
on the LS whereas it increases on the TS.
This has several consequences: 1) As shown with the horizontal arrows, 
the central component (`core')
lags the centroids of pairs of outer components.
2) The leading components become wider than the trailing ones
(see Ahmadi and Gangadhara 2002, and Dyks et al.~2010b 
for observational evidence).
3) Because $\dcut$ is smaller on the LS, 
for decreasing $\nu$ the leading components
departure from the `core' faster than the trailing components.
This is marked with the short thick bars that connect
the centres of fixed-$\nu$ ellipses with the sightline 
path.\footnote{Note that the bars do not mark the observed peak positions
of fixed-$\nu$ components, as was the case in Fig.~\ref{rfm},
because the elliptic shape of the intensity decline (Fig.~\ref{specgrad}b) 
is ignored to maintain the clarity of
Fig.~\ref{aber}. This is a 
technical detail which does not change our conclusions
 qualitatively. 
The faster departure of the LS components
also exists for circular patterns (Fig.~\ref{centroid}), which were not used in
Fig.~\ref{aber} to make it look more realistic.}
Fig.~4 in GG01 presents the observed 
version of this effect.
4) The angular distance $\theta$ from the sky-projected, low-altitude
dipole axis  becomes larger
for the leading components, than for the `corresponding'
components on the TS. This contributes to the 
left-right asymmetry of the profile, because the LS
patterns become more spread on the sky, so the peak flux of a leading
component decreases in comparison to a `corresponding' trailing component. 
This effect should be strongest for fast rotators,
and may be responsible for the ubiquitous weakness of the LS
in main pulses of MSPs (see eg.~Fig.~2 in DRD10; Demorest 2007).
Thus the extension of the LS 
emission over a long stretch of pulse longitude occurs also in the
stream model, along with the simultaneous pile up of emission
on the TS (see.~Dyks et al.~2010b).
5) Since the consecutive ellipses in the trailingmost stream 
are more overlapping with each other, the spectrum is more broadband 
%gradient spectral scale
on the TS. This enhances the asymmetry 
of components'
shifts thereby increasing the centroid shift.\footnote{This can be
understood as follows: Let us assume that a fixed frequency interval, say
between $0.4$ GHz and $1.4$ GHz, corresponds to the transition from 
one elliptic
contour to the adjacent one. For a more broadband spectrum
%smaller $\theta$ spectral gradient
%spectral scale
(more overlapping ellipses on the TS), the angular 
displacement 
(measured along the
beam) of a $0.4$-GHz contour from a $1.4$-GHz contour is smaller. Hence
the shift of the trailing component is smaller than
that of the leading one, for which  
the spectrum is less broadband
%the spectral gradient is larger 
%spectral scale
(less overlapping ellipses).}
Thus, the AR-induced asymmetry does not only generate the shift through
the asymmetry of $\dcut$. It may also 
make the TS spectrum more broadband, 
%decrease the angular spectral gradient on the TS
%spectral scale
increasing the disproportion of the
shifts on the LS and TS.
\footnote{Physically, the stretched angular
distance between the leading component and the dipole axis, 
corresponds to the larger curvature of electron trajectories 
on the LS (see Fig.~2 in Dyks et al.~2010b;
Thomas et al.~2010).}

Because of the spectral gradient asymmetry, the trailingmost component
 should have  
a broader spectrum, ie.~extending over a wider range of $\nu$, 
 than the leadingmost
component. The TS of a main pulse
should then exhibit weaker shape evolution with $\nu$ 
than the trailing side, as observed eg.~for PSR J1012$+$5307
(Dyks et al.~2010b).

\subsubsection{Simple model for the AR-induced shift of centroids}

As in Sect.~\ref{mathacc}, 
we model the intermediate case of Fig.~\ref{specgrad}b
with the circular patterns of Fig.~\ref{specgrad}a. 
The pulse longitude of a component is given by
\begin{equation}
\phi=\phi^\prime + \Delta\phi_{\sss AR} = \theta\sin\phm - k\theta^2,
\label{faza2}
\end{equation}
where $\phi^\prime$ represents the pulse longitude observed for a beam 
with no AR effects included (Fig.~\ref{aber}a, eq.~\ref{faza}),
and the AR shift is equal to
\begin{equation}
\Delta\phi_{\sss AR}
= -2\frac{r}{\rlc} = -\frac{8}{9}\frac{\theta^2}{s^2} = -k\theta^2,
\label{arshift}
\end{equation}  
where $k=8/(9s^2)$, and $s=\thm/(r/\rlc)^{1/2}$ is the footprint parameter
(eg.~Arons 1983). 
In the case of a non-orthogonal
dipole inclination $\alpha$, the first term of eq.~\mref{faza2} 
(but not the second one) should be rescaled roughly by a factor $1/\sin\alpha$
but this effect is ignored because it is symmetric in $\phm$.
The observed pulse longitude of a component changes with decreasing frequency 
at the rate of
\begin{equation}
-\frac{d\phi}{d\nu} = -\frac{d\phi}{d\theta}\dsp = -\dsp\left(\sin\phm 
- 2k\theta\right).
%\label{}
\end{equation}
 As before, the negative derivative is considered to learn how the profiles
evolve with decreasing frequency.
% ($-d\nu=\nu_{\rm low}-\nu_{\rm high}$).
The central component ($\phm=0$) therefore moves leftward at the `spectral 
speed'
\begin{equation}
v_{\rm ctc} = %\left|
-\frac{d\phi_{\rm ctc}}{d\theta} \dsp =  -2k\dspc\theta_{\rm ctc},
\label{core2}
\end{equation}
where $\theta_{\rm ctc}$ is the magnetic colatitude  
of the propagation direction for the point
at which the line of sight
 is passing through the `core'. %Spectral gradients 
The angular scales of spectral variations
for different
components are determined as in Sect.~\ref{mathacc},
ie.~$\dsp=-\dspc$,
$\dsp=-\dspl$, etc. 
For a pair of flanking components at magnetic azimuths $\phm^L$ and 
$\phm^T$ we have $d\phi_i/d\theta = 
\sin\phm^i - 2k\theta_i$, where $i=L$ or $T$, so their leftward 
speeds across a profile are:
\begin{equation}
v_{\sss L} = %\left|
-\frac{d\phi_{\sss L}}{d\theta}\dsp
= \dspl\left[\sin(-|\phm^{\sss L}|) -2k\theta_{\sss L}\right]
\end{equation}
\begin{equation}
v_{\sss T} =
-\frac{d\phi_{\sss T}}{d\theta}\dsp
= \dspt\left[\sin(\phm^{\sss T})-2k\theta_T\right]
\end{equation}
In the case of the symmetric properties 
of the stream system with respect
to the MM ($\phm^{\sss L}=\phm^{\sss T}=\phm^{\rm pair}$, $\theta_{\sss L}=
\theta_{\sss T}=\theta_{\rm pair}$, $\dspl=\dspt=\dspp$), 
the centroid moves leftward at the speed:
\begin{equation}
v_{\rm pair} = \frac{v_{\sss L} + v_{\sss T}}{2} = -2k\dspp\theta_{\rm pair},
\label{vel2}
\end{equation}
where $\theta_{\rm pair}$ is the $\theta$-angle for the locations of the 
sightline-cut  through the flanking streams.
In agreement with the expectations based on Fig.~\ref{aber}, 
for $\dspc=\dspp$ the centroid speed
is larger than  the `core' speed $v_{\rm ctc}$, because 
at the locations probed by the line of sight $\theta_{\rm ctc}
< \theta_{\rm pair}$.
The observed $\nu$-dependent speed at which the centroid 
moves with respect to the
central component is given by:
\begin{eqnarray}
v_{\rm obs} & = & v_{\rm pair} - v_{\rm ctc} = -2k\dspp\theta_{\rm pair}
+2k\dspc\theta_{\rm ctc} = \nonumber\\
&  = & -2k(\dspp\theta_{\rm pair}-\dspc\beta),
\label{vvv}
\end{eqnarray}
Eq.~\mref{vvv} has the required
property of producing larger shift
of centroid for the more peripheral (outer) pairs of components 
(which have larger $\theta$ at the sightline-crossing points).
Moreover, the shift increases with decreasing $\nu$, because 
$v_{\rm obs} \propto \theta$ and $d\theta/d\nu < 0$ %holds
for the pulsars analysed in GG03.
The AR effect  
can thus generate 
all the observed $\nu$-dependent effects exhibited by these pulsars.
In contrast to the accidental shift caused by the beam-system
misalignment, the AR effects displace the centroid in only one direction
(leftward, provided $r$ is smaller for inner components).
The AR shift can therefore be either enhanced or made weaker by the misalignment.
If locations of streams in pulsars are accidental, 
in most cases the central component should lag the centroids.
The opposite arrangement (lag of the centroid) can occur only
for objects in which the misalignment dominates over the AR effects.

The widely-used method of $r$-determination from the `core' lag needs to be
reconsidered in view of the stream interpretation.
If the streams in the corotating pulsar magnetosphere are so symmetric
as shown in Fig.~\ref{aber}a, then, from eq.~\mref{faza2}, the
`core' lag is equal to: 
\begin{eqnarray}
\Delta\phi_{\rm obs} & = & \phi_{\rm pair} -\phi_{\rm ctc}=-k\theta_{\rm pair}^2
+k\theta_{\rm ctc}^2 =\nonumber\\
 & = & -k(\theta_{\rm pair}^2-\beta^2) = -2\left(\frac{r_{\rm pair}}{\rlc}
-\frac{r_{\rm ctc}}{\rlc}\right)
\label{rdeter}
\end{eqnarray}
Thus, the published emission altitudes
that were derived within the conal model may still represent the
altitude differences between the `core' component
and the flanking components. 
This is because it does not matter whether the AR-shifted patches 
of intensity 
are fragments of a single cone or fragments of two symmetric streams. 
What has changed is the interpetation of
the altitude -- it now represents the altitude of that part of a stream
that happened to be probed by our line of sight. The `core' altitude, eg.,
represents the altitude of the sightline-crossing point -- it does not
have to be next to the neutron star surface.
 
It needs to be emhasized that the AR-induced lag of `core' is 
a robust effect which operates also for not perfectly symmetric system of 
streams. In the asymmetric case shown in Fig.~\ref{centroid},
the AR effects would also bend the streams forward.
The components would accordingly move forward in phase, enhancing
the `core' lag that results from the misalignment alone.
In the case of the leftward misalignment of the beam system (negative
$\epsilon$) the AR effects could possibly lead to the `core' lag,
provided they were stronger than the effects of the misalignment,
which occurs roughly for $\epsilon < \theta$. 
The magnitude of the AR shift quickly increases with $\theta$ so the 
AR-induced `core' lag is likely to appear also for
systems of streams which are not fully symmetric with respect to the MM.

However, the fact that the \emph{magnitude} of 
the centroid shift can be affected 
by \emph{any} asymmetry of the type shown in Fig.~\ref{centroid},
makes the altitudes obtained with the AR-lag method 
less reliable. Direct use of 
eq.~\mref{rdeter} is risky, because it assumes
perfect  symmetry of the stream system and ignores spectral differences
between the streams  
(eg.~one may have $\theta_L(\nu) \ne \theta_T(\nu)$ for
the flanking streams). 
 In the case of the intrinsically asymmetric beam structure,
results of Sect.~\ref{accidental} need to be convolved with the AR lag.

Therefore, whereas the AR-induced lag can be observed also for the
intrinsically non-symmetric
geometry, eq.~\mref{rdeter} is valid only for the fully symmetric case. 

We conclude that the AR effects working within streams
with spatial spectral gradient
(spatial RFM) explain both the broken symmetry of profiles,
as well as the increasing strength of this effect with decreasing
frequency. We emphasize that in the stream model
the low altitude of the `core' is not an ad-hoc assumption made to produce
the observed centroid shift. In the stream model the low-$r$ location 
of `core' naturally results from the dipolar geometry of streams,
which near the dipole axis (small $\theta$) occupy small $r$.
In the case of the inner components (associated with more meridional
streams) the line of sight cuts through the streams
at a smaller angle $\theta$ and smaller $r$, 
provided the streams have similar $s$.

We have discussed only the simplest case of identical streams 
located at different azimuth $\phm$. A much larger variety of streams 
(with different $s$, $r$,
$\thm$, $\dsp$, and azimuthal widths and separations) 
apparently exist in magnetospheres of real pulsars. 
This would explain the diversity of observed properties of pulsar profiles.

\begin{figure}
	\centering
	\includegraphics[width=0.5\textwidth]{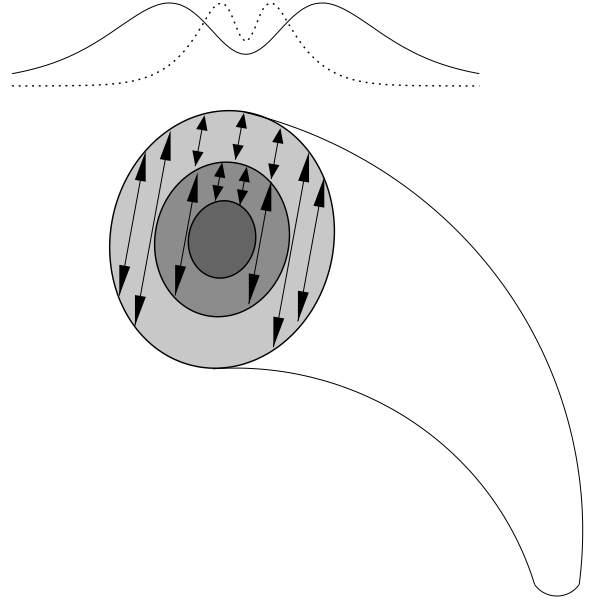}
\caption{The cross-section through a magnetospheric stream, exposing
the transverse decline of plasma density (grey contours). 
The arrows show that the sightline-integrated emissivity
is larger at the azimuthal sides of the stream. Moreover, the emission 
from the bottom part of the stream can be obscured 
by the central high-density region. The resulting pulse profiles 
are shown on top. The dotted one is for a larger $\nu$.
}
\label{robak}
\end{figure}

\section{Obscuration effects and bifurcated emission components}
\label{obscur}

In the cone/core scenario the radio emission was believed to originate from 
well-separated and narrow intervals of altitude 
and footprint parameter. This was not likely to lead to obscuration effects.

In the case of elongated streams, the geometry that leads to eclipses
of lower-$r$ emission can appear in at least two cases.
First, this happens when a laterally extended, quasi-uniform emission 
is blocked from our view by a dense non-emitting stream.
This case was shown to produce deep double notches in the profiles
of both normal and millisecond pulsars (DRD10; DR12; McLaughlin \& Rankin
2004).

Second, when the central high-density region of the stream extends sufficiently far
in the longitudinal direction, the bottom parts of the stream, located on its
concave side, can also be eclipsed (ie.~the radiation from 
this region can be scattered or absorbed and dissipated).
This can be seen in Fig.~\ref{visi}, in which the lowermost `ray', emitted
near the bottom of the lower emission region, must 
cross the full thickness of the region before reaching the outer
low-density regions. Depending on the transverse gradient of plasma
density in the stream, and on the observed frequency, the ray can be 
obscured. This can diminish the flux observed at the centre of a component,
producing a bifurcated emission component (BEC), similar to those observed
in PSR J0437$-$4715 and J1012$+$5307.
Moreover, the tubular shape of
transverse density pattern  (Fig.~\ref{robak}) 
should additionally enhance the flux in the azimuthally-outer parts of the
associated component. This is because the radio emission 
occurs within the planes of fixed magnetic azimuth $\phm$ (if the extent of
the $1/\gamma$
microbeam is negligible). Crossections of the stream with these 
fixed $\phm$ planes are shown with the arrows in Fig.~\ref{robak}.
Different zones of equal
density are shown with different grades of grey. 
Since the radio emission 
occurs in the planes containing the arrows and the dipole axis, 
the largest emissivity per steradian
corresponds to the azimuthal sides of the stream.
Thus a `limb brightening' can be expected. 
Emission at a larger frequency originates from the inner regions
 of the stream (the inner ring in Fig.~\ref{robak}) which implies
the narrowing of BECs with frequency, consistent with the 
observations of PSR J0437$-$4715 and J1012$+$5307.
Thus, the nature of BECs may be macroscopic, and not related
to the microphysical (elementary) emission beam.
This would solve the problem of large energetic requirements
implied by the microscopic interpretation (Dyks \& Rudak 2013; 
Gil \& Melikidze 2010). However, it is less clear why 
the BECs should have so high symmetry as observed, 
because rotation tends to remove the stream
away from the path of a to-be-eclipsed ray.
It is also important to note that in the superstrong magnetic field 
only the radiation in the ordinary 
polarisation mode would be eclipsed. The extraordinary polarisation mode,
however, is not emitted within the $\phm$-plane anyway (DRD10; Gil et
al.~2004).
The mode-dependent obscuration would introduce imbalance between
the observed contribution of orthogonal polarisation modes, which should
change with the distance from a BEC's centre.

It is also possible that both factors
(micro- and macroscopic) shape the BECs.
Regardless of this issue, the general interpretation of the BECs 
in MSPs remains similar to the one described in DRD10 and DR12: 
they are caused by the broad-band,
locally-bidirectional\footnote{`Bidirectional' means the emission 
into the lobes
of the bifurcated beam shown in Fig.~1 of DR12. It does not mean the
emission in two opposite directions.}
radio emission,
and are observed when our line of sight passes through streams
in pulsar magnetosphere. 
The only difference is that in the macroscopic case, the quasi-local
bifurcation is attributed to the opacity of a segment of a narrow stream.

With the possibility of the macroscopically-generated, locally-bidirectional 
beams, 
one needs to ask if double notches can also be created by obscuration
of these macroscopic bifurcated beams.  
This does not seem to be possible, 
because the emitter associated with the phenomenon 
of double notches (see Fig.~12 in DRD10) has to be extended in two
dimensions (azimuth and latitude) to be consistent with the
quasi-isotropic radio emission in which double notches are observed.
The emitter does not have the geometry of the stream
shown in Fig.~\ref{robak}.
Moreover, the notches are 
well-fitted with the curvature radiation microbeam (Fig.~6 in DR12),
have different shapes than BECs (Sect.~3.3 in DR12),  
and all of them are narrower than the problematic BEC of PSR J1012$+$5307
(see Fig.~4 in Dyks \& Rudak 2013). 
Therefore, the microscopic nature of double notches 
continues to present a promising model.

\section{Conclusions}

We have shown that the main frequency-dependent phenomena exhibited
by radio pulsar profiles: the apparent RFM and the frequency-dependent
centroid shift can be naturally explained within the stream model 
-- the model which
assumes fan-shaped radiation beams produced by streams
flowing in pulsar magnetosphere.

By considering the spatial spectral gradient along the stream, 
we have shown that even some detailed aspects of the phenomena
can be effortlessly reproduced by the stream model.
These include the
weaker apparent RFM of the inner `conal' pair, as well as the 
increasingly large centroid shift at lower frequencies.
The weak spreading of inner pair components is a trivial consequence
of more orthogonal stream cutting.
In the stream model, the shift of centroids does not require
the complicated and arbitrary positioning of the core, inner
cone and outer cone emission at disparate altitudes and magnetic 
colatitudes. Instead, the altitude differences result naturally
from the geometry of the cut itself.

Surprisingly the shifts of centroids of both signs
can appear even with no altitude differences 
(and no AR affects) associated with the fan beam.  
However, only the backward misalignment of the beam system
has the properties consistent with
pulsars described in GG01 and GG03.
Whenever both the shift-inducing conditions exist, ie.~when
the symmetric system of beams is misaligned and emission altitude increases
towards the periphery of the beam, then the AR shift can be enhanced,
diminished or reversed by the effects of the misalignment.
In contrast to the misalignment shift, the AR shift
always provides 
the type (sign) of the asymmetry and the frequency behaviour
which is observed for pulsars described by GG03.
This naturally results from the geometry of streams which 
in the dipolar field occupy low altitudes at small $\theta$.

In the case of the stream model, even for identical
locations of the streams on their corresponding planes
of fixed magnetic azimuth (same $\Delta s$, $\Delta r$),
the approximate left-right symmetry of profile properties
is likely to appear because smaller altitudes are sampled for inner 
(more meridional) streams. The `core' component is created by the sightline
cut through a near-meridional stream whereby the lowest altitudes 
are probed.
The altitude differences required to produce the centroid shift
ensue naturally, because the line of sight picks up
a lower-altitude emission in the middle of the profile, ie.~no radial
displacements need to be introduced between different streams
to generate the AR lag of the central component (though in general such
differences in $r$ are not excluded).
The stream model can also explain some differences between the
frequency-dependent behaviour of normal pulsars and MSPs.

It has been shown that by introducing the azimuthal non-uniformity
to the pulsar beam one can avoid several
problems which remain challenging for the conal model.
 In particular, the problem of origin
of the `core' component in the conal model disappears, with the role of the
physical core region taken over by the central stream. 
Interestingly, the observed `core' emission
is still found to originate from the lowest altitudes 
and this is probably why its properties are sometimes observed to be
different from properties of the outer components.

Similar effects (decrease of $r$ towards the centre of the profile)
are likely responsible for the apparent symmetry of
fluctuation spectra with respect to the centre of profiles
(Backer 1970; Maan \& Deshpande 2014). 
The origin of subpulse drift (eg.~Weltevrede et al.~2007;
Ruderman \& Sutherland 1975), 
the apparent `carousel circulation' (Deshpande \& Rankin 2001)
and other single-pulse effects certainly deserve a careful 
study within the stream model with azimuthally-structured average emission 
pattern. The new type of radio beam may also
impact the population studies (Story et al.~2007) and the multiwavelength
lightcurve fitting (Pierbattista et al.~2014; Venter et al.~2012;
Romani \& Watters 2010). 
The elongated beams
also provide the flexibility that is needed to interpret profiles of
difficult objects such as the Crab pulsar (Moffett \& Hankins 1996)  
or PSR B1821$-$24 (Johnson et al.~2013; Knight et al.~2006).

The azimuthal structure calls for explanation itself. Results of this paper
suggest that more efforts should be put into the understanding 
of the azimuthal
nonuniformity of pulsar magnetosphere, 
instead of the radially-separated emission rings, which
seemed to be justified by the `core'-lag effect, or by the general
applicability of the conal classification scheme.

The apparent RFM and the centroid shift
were providing considerable support for the conal beam model. 
However, we have shown that these conal interpretations
can be successfully replaced with the stream-based reasoning. 
Furthermore the stream model is able to explain even such
`peculiarities' as the bifurcated components and double notches, 
which makes it a very successfull tool to interpret
radio pulsar beams.

The evidence for the stream model (or fan-beam model)
is mounting steadily.
As recently shown by Wang et al.~(2014) the fan beam model
is more successful in reproducing the observed  statistics of profiles' 
width and flux than the conal model.
In the upcoming paper (Pierbattista \& Dyks, in preparation) 
we show that the
statistics of component locations within the profiles
 with 4 and 5 components (Q and M class)
is inconsistent with the nested cone model.
Teixeira et al.~(2014) show that probably the most symmetric
example of the Q-type profile (J0631$+$1036, Zepka et al.~1996) 
is best explained by the multiple-stream
model, which avoids several problems of the conal interpretation.
All this development is additionally supported
by the recent beam maps of precessing pulsars, including the 
feebly constrained but fairly suggestive map of J1906$+$0746.
The difficulties in reproducing the 
beam maps of other precessing pulsars with the use of conal beams,
are likely caused by application of inappropriate beam topology.

The stream model is then supported by variety of arguments,
it is able to explain several pulsar features, and avoids
several problems of the conal model.
With no reserve we therefore advance the view that pulsars possess
fan beams emitted by outflowing plasma streams, 
whereas the nested cone/core
beams may well not exist.

\section*{Acknowledgments}
%\noindent {\bf Acknowledgments:}
We thank Patrick Weltevrede for a careful review.
The work presented in this paper was funded by the Polish National 
Science Centre grant DEC-2011/02/A/ST9/00256. 

\section*{REFERENCES}

%\begin{thebibliography}{99}
%\bibliographystyle{plainnat}

%\bibitem{amd97} Ables, J.G., McConnell, D., Deshpande, A.A., \& Vivekanand, M.
%1997, ApJ, 475, L33
%\bibitem{ag02} 
\noindent Ahmadi P., Gangadhara R.T., 2002, ApJ, 566, 365\\
%\bibitem{a783} 
Arons J., 1983, ApJ, 266, 215\\
%\bibitem{as79} 
Arons J., \& Scharlemann E.T., 1979, 231, 854\\
%\bibitem{bs2010} Bai X.-N., Spitkovsky A., 2010, ApJ, 715, 1282
%\bibitem{b70} 
Backer, D.C., 1970, Nature, 228, 752\\
%\bibitem{bkh82} Backer, D. C., Kulkarni, S. R., Heiles, C., Davis, M. M. \& Goss, W. M., 1982, Nature, 300, 615
%\bibitem{b86} Barnard, J.J. 1986, ApJ, 303, 280
%\bibitem{ba86} Barnard,J.J. \& Arons, J., 1986, ApJ, 302, 138
%\bibitem{bcc2003} Bhat N.D.R., Cordes J.M., \& Chatterjee S. 2003,
%ApJ, 584, 782
%\bibitem{bggs} Bhattacharyya B., Gupta Y., Gil J., Sendyk M., 2007, MNRAS,
%377, L10
%\bibitem{bcw91} Blaskiewicz M., Cordes J.M., Wasserman I., 1991, ApJ, 370, 643
%\bibitem{brr2000} Braje, T.M., Romani, R.W., \& Rauch, K.P. 2000, ApJ, 531, 447
%\bibitem{bpm05} 
Burgay M., Possenti A., Manchester R.N., Kramer M.,
McLaughlin M.A., et al.~2005, ApJ, 624, L113\\
%\bibitem{bb80} Buschauer R., Benford G., 1980, MNRAS, 190, 945
%\bibitem{cjd04} Cairns, I.H., Johnston, S., \& Das, P. 2004, MNRAS, 353, 270
%\bibitem{crh06} Camilo, F., Ransom, S. M., Halpern, J. P., Reynolds, J.,
%Helfand, D. J., et al., 2006, Nature, 442, 892
%\bibitem{ccr07} Camilo, F., Cognard, I., Ransom, S. M., Halpern, J. P., Reynolds, J., 
%et al., 2007, ApJ, submitted (astro-ph/0610685)
%\bibitem{cr77} Cheng, A. F., \& Ruderman, M. 1977, ApJ, 216, 865
%\bibitem{cr79} Cheng, A. F., \& Ruderman, M. 1979, ApJ, 229, 348
%\bibitem{chr86} Cheng K. S., Ho C., \& Ruderman M. A., 1986, ApJ, 300, 500
%\bibitem{crz00} Cheng K. S., Ruderman M. A., \& Zhang L., 2000, ApJ, 537, 964
%\bibitem{cw08} 
Clifton T., \& Weisberg J.M., 2008, ApJ, 679, 687\\
%\bibitem{c75} Cordes J. M., 1975, ApJ, 195, 193
%\bibitem{c78} 
Cordes J. M., 1978, ApJ, 222, 1006\\
%\bibitem{cwh90} Cordes, J. M., Weisberg, J.M., \& Hankins, T.H. 1990, AJ 100.2,
% 1882
%\bibitem{dr07} D'Angelo, C., \&  Rafikov, R.R. 2007, Phys.~Rev.~D, 75, 042002
%\bibitem{dh83} Daugherty J.K., Harding A.K., 1983, ApJ 273, 761
%\bi de Jager, O.C., 2002, BASI, 30, 85 
%\bibitem{d2007} 
Demorest P., 2007, PHD Thesis, Univ.~of California, Berkeley\\
%\bibitem{dr2001} 
Deshpande A.A., \& Rankin J.M., 2001, 322, 438\\
%\bibitem{dkc2013} 
Desvignes, G., Kramer, M., Cognard, I., Kasian, L.,
van Leeuwen, J., Stairs, I., \& Theureau G. 2013, 
Proc.~IAU Symposium No.~291, ed.~J.~van Leeuwen, 199\\
%\bibitem{d55} Deutsch, A.J. 1955, Ann.~d'Astrophys., 18, 1
%\bibitem{d08} Dyks J., 2008, MNRAS, 391, 859
%\bibitem{dh04} Dyks J., \& Harding A. K., 2004, ApJ 614, 869
%\bi Dyks, J., Rudak, B., 2000, MNRAS, 319, 477 
%\bibitem{dr02} Dyks, J., \&  Rudak, B. 2002, A\&A, 393, 511
%\bibitem{dr03} Dyks, J., \&  Rudak, B. 2003, ApJ, 598, 1201
%\bibitem{dfs05} Dyks J., Fr{\c a}ckowiak M., S{\l}owikowska A., Rudak B., 
%\& Zhang B., 2005, ApJ 633, 1101
%\bibitem{dhr04a} Dyks J., Harding A. K., \& Rudak B., 2004, ApJ 606, 1125
%\bibitem{dr12} 
Dyks J., \& Rudak B., 2012, MNRAS, 420, 3403 (DR12)\\
%\bibitem{dr13} 
Dyks J., \& Rudak B., 2013, MNRAS, 434, 3061\\
%\bibitem{drd10} 
Dyks J., Rudak B., \&  Demorest P., 2010a, MNRAS, 401, 1781 
(DRD10)\\
%   (DRD10)
%\bibitem{drh04b} 
Dyks J., Rudak B., \& Harding A. K., 2004, ApJ 607, 939\\ %(DRH)
%\bibitem{drr07} Dyks J., Rudak B., \& Rankin J. M., 2007, A\&A, 465, 981 
%(DRR07)
%\bibitem{dwd10} 
Dyks J., Wright G.A.E., \&  Demorest P., 2010b, MNRAS, 
405, 509\\
%\bibitem{dzg05} Dyks, J., Zhang, B., \& Gil, J. 2005b, ApJ, 626, L45
%\bibitem{e04} Edwards R.T., 2004, A\&A, 426, 677
%\bibitem{es04} Edwards R.T., \& Stappers, B.W., 2004, A\&A, 421, 681
%\bibitem{esv03} Edwards, R.T., Stappers, B.W., \& van Leeuwen, A.G.J. 2003,
%  A\&A, 402, 321
%\bibitem{e73} Epstein, R. I., 1973, ApJ, 183, 593
%\bibitem{ew01} Everett J.E., \& Weisberg J.M., 2001, ApJ 553, 341
%\bibitem{fw82} Fowler, L.A., \& Wright, G.A.E. 1982, A\&A, 109, 279 
%\bibitem{fl04} Fussell D., \& Luo Q., 2004, MNRAS, 349, 1019
%\bibitem{fl04} Fussell D., Luo Q., \& Melrose D.B., 2003, MNRAS, 343, 1248
%\bibitem{gg01}
Gangadhara R. T., \& Gupta Y., 2001, ApJ, 555, 31 (GG01)\\
%\bibitem{gk93} 
Gil J., \& Kijak J., 1993, A\&A, 273, 563\\
%\bibitem{gm10} 
Gil J.A., \& Melikidze G.I., 2010, astro-ph/1005.0678\\
%\bibitem{gk97} Gil, J., \& Krawczyk, A. 1997, MNRAS, 285, 561
%\bibitem{gs90} Gil J., \& Snakowski J.K., 1990, A\&A, 234, 237
%\bibitem{glm04} 
Gil J., Lyubarsky Y., \& Melikidze G.I., 2004, ApJ, 600, 872\\
%\bibitem{gh94} Gonthier, P.L., \& Harding, A.K. 1994, ApJ, 425, 767
%\bibitem{gl98} Gould, D.M., \& Lyne, A.G. 1998, MNRAS, 301, 235
%\bibitem{gkj2013} Guillemot, L., Kramer M., Johnson, T.J.,
% Craig, H.A., Romani, R.W., Venter C., Harding, A.K., Ferdman, R.D., 
% et al. 2013, ApJ, 768, 169
%\bibitem{gg03} 
Gupta, Y., \& Gangadhara, R.T. 2003, ApJ, 584, 418 (GG03)\\
%\bibitem{hpy07} Haensel P., Potekhin A. Y., Yakovlev D.G., 2007,
% Neutron Stars 1: Equation of State and Structure, Astrophysics and space
% science library, Vol.~326, New York, Springer
%\bibitem{hc81} Hankins, T.H., \& Cordes, J.M. 1981, ApJ, 249, 241
%\bibitem{hr08} 
Hankins, T.H., \& Rankin, J.M., 2010, ApJ, 139, 168\\
%\bibitem{hm11} Harding A.K., \& Muslimov A.G., 2011, ApJ, 726, L10
%\bibitem{hum05} Harding, A.K., Usov V., Muslimov A.G., 2005, ApJ 622, 531
%\bibitem{hr08} Herfindal J., \& Rankin J.M., 2008, MNRAS, submitted
%\bibitem{hrs06} Hessels, J. W. T., Ransom, S. M., Stairs, I. H., Freire, P. C. C., Kaspi, V. M. \& Camilo, F., 2006, Science, 311, 1901
%\bibitem{ha01} Hibschman, J.A., \& Arons, J. 2001, ApJ, 546, 382 (HA)
%\bibitem{h07} Hirotani K., 2007, ApJ, 662, 1173
%\bibitem{hhs03} Hirotani, K., Harding, A.K., \& Shibata, S. 2003, ApJ, 591, 334
%\bibitem{j75} Jackson J.D., 1975, ``Classical Electrodynamics", John Wiley
% \& Sons Inc, New York
%\bibitem{jak98} Jenet, F.A., Anderson, S.B., Kaspi, V.M., et al. 1998, ApJ,
% 498, 365
%\bibitem{jgk2013} 
Johnson, T.J., Guillemot, L., 
Kerr, M., Cognard, I., Ray, P.S., Wolff, M.T., B\'egin, S., 
Janssen, G.H. 2013, ApJ, 778, 106\\
%\bibitem{jr02} Johnston, S., \& Romani, R.W. 2002, MNRAS, 332, 109
%\bibitem{jkmg08} Johnston, S., Karastergiou, A., Mitra, D., \& Gupta, Y. 2008,
%MRAS, in press (astro-ph/0804.3838)
%\bibitem{jw06} Johnston S., Weisberg J.M., 2006, MNRAS, 368, 1856
%\bibitem{kl10} Kaganovich A., \& Lyubarsky Y., 2010, ApJ, 721, 1164
%\bi Kanbach, G., 1999, Astrophysical Letters \& Communications, 38, 17
%\bibitem{karas2009} Karastergiou, A. 2009, MNRAS, 392, 60
%\bibitem{kj07} 
Karastergiou, A., \& Johnston, S. 2007, MNRAS, 380, 1678\\
%\bibitem{kg97} Kijak, J., \& Gil, J. 1997, MNRAS, 288, 631
%\bibitem{kbm2006} 
Knight, H.S., Bailes, M., Manchester, R.N., 
Ord, S.M. 2006, ApJ, 653, 580\\
% \bibitem{k70} 
Komesaroff, M.M. 1970, Nature, 225, 612\\
%\bibitem{k81} Konopinski E.J., 1981, ``Electromagnetic fields and
%relativistic particles", McGraw-Hill Book Company, New York
%\bibitem{k94} 
Kramer M., Wielebinski R., Jessner A., Gil J.A., \& Seiradakis
J.H., 1994, A\&AS, 107, 515\\
%\bibitem{k08} 
Kramer M., 1998, ApJ, 509, 856\\
%\bibitem{klo06} Kramer, M., Lyne, A.G., O'Brien, J.T., et al.~2006, Science,
%  312, 549
%\bibitem{kxl98} 
Kramer M., Xilouris K. M., Lorimer D., Doroshenko O.,
Jessner A., Wielebinski R., Wolszczan A., Camilo, F. 1998, ApJ, 501, 270\\
%\bibitem{kll99} 
Kramer M., Lange C., Lorimer D., Backer D.C., 
Xilouris K. M., Jessner A., \& Wielebinski R. 1999, ApJ, 526, 957\\
%\bibitem{kx00} Kramer M., \& Xilouris K. M., 2000, Pulsar Astronomy -- 2000
%and Beyond, ASP Conference Series,
%Vol.~202, eds.~M. Kramer, N. Wex, \& R.~Wielebi\'nski
%\bibitem{kd83} Krishnamohan, S., \& Downs, G.S. 1983, ApJ, 265, 372 (KD83) 
%\bibitem{klj98} Kunzl T., Lesch H., Jessner A., von Hoensbroech, 1998, ApJ
%  505, L139
%\bibitem{ki93} Kuzmin, A.D., \& Izvekova, V.A. 1993, MNRAS, 260, 724
%  \bibitem{kis98} Kuzmin, A.D., Izvekova, V.A., Shitov, Yu.P., et al. 1998,
%  A\&AS, 127, 355
%\bibitem{kl01} 
Kuzmin, A.D., \& Losovsky, B.Ya., 2001, A\&A, 368, 230\\
%\bibitem{lcc01} Lai, D., Chernoff, D.F., \& Cordes, J.M. 2001, ApJ, 549, 1111
%  \bibitem{lkw98} Lange, Ch., Kramer, M., Wielebinski, R., \& Jessner, A. 1998
%  A\&A, 332, 111
%\bibitem{lmj05} Levinson, A., Melrose, D., Judge, A., \& Luo, Q. 2005, 
%   ApJ, 631, 456
%\bibitem{lh03} Li X.H., \& Han J.L. 2003, A\&A, 410, 253 
%\bibitem{lk05}Lorimer D.R., \& Kramer, M., 2005, {\it "Handbook of pulsar
%astronomy"}, Cambridge University Press, Cambridge (LK05)
%\bibitem{lsf06} 
Lorimer D.R., Stairs I.H., Freire P.C., Cordes J.M., 
Camilo F., et al., 2006, ApJ, 640, 428\\
%\bibitem{lzb} Lommen A.N., Zepka A., Backer D.C., McLaughlin M., Cordes
%J.M., et al., 2000, ApJ, 545, 1007
%\bibitem{lm92} Luo, Q., \& Melrose, D. B. 1992, MNRAS, 258, 616 
%\bibitem{lm95} Luo, Q., \& Melrose, D. B. 1995, MNRAS, 276, 372 
%  \bibitem{lm06} Luo, Q., \& Melrose, D. B. 2006, MNRAS, 371, 1395 
%\bibitem{ml88} 
Lyne A.G., \& Manchester, R. N., 1988, MNRAS, 
234, 477\\
%\bibitem{l02} Lyubarsky Y.E., 2002, 
%  in Becker W., Lesch H., Tr\"umper J., eds., MPE Report 278
%  Proc. 270th WE-Heraeus Seminar, Max-Planck-Institut f\"ur extraterrestrische
%  Physik, Garching bei M\"unchen, p. 230
%  \bibitem{lp98} Lyubarsky, Y.E., \& Petrova, S.A. 1998, A\&A, 337, 433
%\bibitem{md2014} 
Maan, Y., \& Deshpande, A.D., 2014, ApJ, accepted
 (astro-ph/1407.6368v1)\\
%\bibitem{mgj1994} 
Malofeev, V.M., Gil, J. A., Jessner, A., Malov, I.F.,
Seiradakis, J.H., Sieber, W., Wielebinski, R. 1994, A\&A, 285, 201\\
%  \bibitem{mm01} Malov, I. F., \& Machabeli, G. Z. 2001, ApJ, 554, 587
%  \bibitem{mh04} Manchester, R.N., Han, J.L. 2004, ApJ, 609, 354
%  \bibitem{mht05} Manchester, R.N., Hobbs, G.B., Teoh, A., \& Hobbs, M.
%  2005, Astron. J., 129, 1993
%\bibitem{m2012} 
Manchester, R.N., 2012, Proc.~Electromagnetic
Radiation from Pulsars and Magnetars, eds.~W. Lewandowski, O.~Maron,
\& J. Kijak, San Francisco, ASP Conference Series 466, 61\\
%\bibitem{mks2010} 
Manchester, R. N., Kramer, M., 
Stairs, I. H., Burgay, M., Camilo, F., Hobbs, G. B., Lorimer, D. R.,
 Lyne, A. G., et al.~2010, ApJ, 710, 1694\\
%\bibitem{mkk2000} 
Maron, O., Kijak, J., \& Wielebinski R., 2000, A\&ASS,
147, 195\\
%\bibitem{mab96} McConnell D., Ables J. G., Bailes M., Erickson W. C.,
%1996, MNRAS, 280, 331
%\bibitem{mr04} 
McLaughlin M. A., \& Rankin J. M., 2004, MNRAS, 351, 808\\
%\bibitem{m06} Melrose D.B., 2006, Chin.~J.~Astron.~Astrophys., Vol.~6, 
%  Suppl.~2, 74
%  \bibitem{m78} Melrose, D.B. 1978, ApJ, 225, 557
%  \bibitem{m79} Melrose, D.B. 1979, Austr.~J.~Phys., 32, 61
%  \bibitem{m00} Melrose, D.B. 2000, ASP Conference Series, 202, 721
%\bibitem{m1987} 
Michel, F.C., 1987, ApJ, 322, 822\\
%\bibitem{mr02} 
Mitra D., \& Rankin J. M., 2002, ApJ, 577, 322\\
%\bibitem{ms04} Mitra, D., \& Seiradakis, J.H. 2004, Proc.~of the 6th 
%    Astrononomical Conference, ed.~Laskarides, P., Editing Office of 
%    the Univ.~of Athens, p.~205 (astro-ph/0401335)
%\bibitem{mr02} Mitra D., Rankin J.M., \& Gupta Y., 2007, MNRAS, 379, 932
%\bibitem{msr04} Mitra D., Sarala S., \& Rankin J. M., 2004, preprint
%\bibitem{mh96} 
Moffett D.A., \& Hankins, T.H., 1996, ApJ, 468, 779\\
%\bibitem{m83} Morini M., 1983, MNRAS, 202, 495
%\bibitem{mh03} Muslimov A. G., \& Harding A.K., 2003, ApJ, 588, 430
%\bibitem{nv83} Narayan R., \& Vivekanand M., 1983, ApJ, 274, 771
%\bibitem{mns97} 
Navarro J., Manchester R. N., Sandhu J. S., 
Kulkarni S.R., Bailes M., 1997, ApJ, 486, 1019\\
%\bibitem{n95} Nicastro L., Lyne A.G., Lorimer D.R., et al., 1995, MNRAS,
% 273, L68
%\bibitem{n03} Nowakowski, L. 2003, Arecibo Newsletter, 36, 8
%\bibitem{pmk2010} 
Perera, B. B. P., McLaughlin, M. A., Kramer, M., 
Stairs, I. H., Ferdman, R. D., Freire, P. C. C., Possenti, A., Breton, R.
P., et al.~2010, ApJ, 721, 1193 \\
%\bibitem{pl85} Perry, T.E., \& Lyne, A. G. 1985, MNRAS, 212, 489 (PL85)
%\bibitem{p00} Petrova S. A. 2000, A\&A, 360, 592
%\bibitem{p02} Petrova, S. A. 2002, MNRAS, 336, 774
%\bibitem{p08} Petrova, S. A., 2008, MNRAS, 384, L1
%\bibitem{pl00} Petrova S. A., \& Lyubarsky Y.E., 2000, A\&A, 355, 1168
% \bibitem{p90} Phillips, J. A. 1990, ApJ, 361, L57
%\bibitem{phg2014}
Pierbattista, M., Harding, A. K., Grenier, I. A., Johnson,
T. J., Caraveo, P. A., Kerr, M., Gonthier, P. L., 2014, A\&A, submitted
(astro-ph/1403.3849)\\
%\bibitem{ptv92} Press W.H., Teukolsky S.A., Vetterling W.T., \& Flannery
%B.P., 1992, ``Numerical Recipes in Fortran", Cambridge University Press, Cambridge
%\bibitem{rc69} Radhakrishnan V., \& Cooke D.J. 1969, 
%  Astrophys.~Lett., 3, 225 (RC69)
%\bibitem{rd01} Radhakrishnan, V., \& Deshpande, A.A. 2001, A\&A, 379, 551
%\bibitem{r83} 
Rankin J.M., 1983, ApJ, 274, 333\\
%\bibitem{r90} 
Rankin J.M., 1990, ApJ, 352, 247\\
%\bibitem{r93} 
Rankin J.M., 1993, ApJ, 405, 285\\
%\bibitem{rr03} Rankin J.M., \& Ramachandran R., 2003, ApJ, 590, 411
%\bibitem{rr97} 
Rankin J.M., \& Rathnasree N., 1997, J. Astrophys. Astron., 18, 91\\
%\bibitem{rrv06} Rankin J.M., Ramachandran R., van Leeuwen J., \&
%Suleymanova S.A., 2006, A\&A, 455, 215
%\bibitem{rrw06} Rankin J. M., Rodriguez C., \& Wright G. A. E., 2006, MNRAS,
%370, 673
%\bibitem{rw08} Rankin J. M., \& Wright G. A. E., 2008, MNRAS, 385, 1923
%\bibitem{rwb2013} Rankin J. M., Wright G. A. E., \& Brown A.M. 2013, 
%   MNRAS, 433, 445
%\bibitem{rwr05} Redman S.L., Wright, G.A.E., \& Rankin, J.M., 2005, MNRAS,
%357, 859
%\bibitem{rw2010} 
Romani R.W., \& Watters K.P., 2010, ApJ, 714, 810\\
%\bibitem{ry95} Romani R.W., \& Yadigaroglu I.-A., 1995, ApJ, 438, 314
%\bibitem{r95} Rowe, E.T. 1995, A\&A, 296, 275
%\bibitem{rr94} Rudak, B., \& Ritter, H., 1994, MNRAS, 267, 513 
%\bibitem{r91} Ruderman M.A., 1991, ApJ, 366, 261
%\bibitem{rs75} 
Ruderman M.A., Sutherland P.G. 1975, ApJ 196, 51\\
%\bibitem{rl79} Rybicki G.P., Lightman A.P., 1979, ``Radiative processes
%    in Astrophysics", Wiley-Interscience, New York
%\bibitem{srk02} Schopper R., Ruhl H., Kunzl T.A., Lesch H., 2002, MPE Report
%      278, 193
%\bibitem{s83} Shitov, Yu.P. 1983, Soviet.~Astron., 27, 314
%\bibitem{skks} S{\l}owikowska A., Kanbach G., Kramer M., \& Stefanescu A.,
%2009, MNRAS, submitted
%\bibitem{s86} Smith F. G., 1986, MNRAS, 219, 729
%\bibitem{smk05} Smits J.M., Mitra D., \& Kuijpers J., 2005, A\&A, 440, 683
%\bibitem{s08} Spitkovsky, A. 2008, AIP Conference Proceedings, 983, 20
%\bibitem{sr05} Srostlik Z., \& Rankin J.M., 2006, MNRAS, 362, 1121
%\bibitem{stc99} Stairs I.H., Thorsett S.E., Camilo F., 1999, ApJSS, 123, 627
%\bibitem{sgh2007} 
Story S.A., Gonthier P.L., Harding A.K., 2007, ApJ, 671, 713\\
%\bibitem{s71} Sturrock, P.A. 1971, ApJ, 164, 529
%\bi Thompson, D.J., 2001, AIP Conference Proceedings, 558, 103
%\bibitem{tsh06} Takata, J., Shibata, S., Hirotani, K., \& Chang, H.-K.
%       2006, MNRAS, 366, 1310
%\bibitem{trw} Teixeira M. M., Rankin J. M., \& Wright G. A. E., 
%Proc.~of Zielona G\'ora conference, eds.: W.~Lewandowski,
%O.~Maron, \& J.~Kijak, ASP Conference Series,
%466, 117
%\bibitem{teix_two} 
Teixeira M. M., Rankin J. M., Wright G. A. E., \& Dyks 
J., 2014, MNRAS, submitted\\
%\bibitem{tg2005} Thomas, R.M.C., \& Gangadhara, R.T. 2005, A\&A, 437, 537
%\bibitem{tgg2010} 
Thomas, R.M.C., Gupta, Y., \& Gangadhara, R.T. 2010, MNRAS, 
406, 1029\\
%\bibitem{vjh2012} 
Venter C., Johnson T.J., Harding A.K., 2012, ApJ, 744, 34\\
%\bibitem{vx97} von Hoensbroech, A., \& Xilouris, K.M. 1997, A\&A, 324, 981
%\bibitem{wtc11} Wang Y., Takata J., \& Cheng K.S., 2011, MNRAS, 414, 2664
%\bibitem{wmj07} Wang N., Manchester R. N., \& Johnston S., 2007, MNRAS, 377,
%1383
%\bibitem{wrhz98} Wang, F.Y.-H., Ruderman, M., Halpern, J.P., \& Zhu, T. 1998,
%                       ApJ, 438, 314
%\bibitem{wlh2010} 
Wang P.F., Lai D., \& Han J.L., 2010, MNRAS, 403, 569\\
%\bibitem{wwh2012} 
Wang P.F., Wang C., \& Han J.L., 2012, MNRAS, 423, 2464\\
%\bibitem{wpz2014} 
Wang, H.G., Pi, F.P., Zheng, X.P., Deng, C.L.,
Wen, S.Q., Ye, F., Guan, K.Y., Liu, Y. \& Xu, L.Q, 2014, 
ApJ, 789, 73\\
%\bibitem{wt02} 
Weisberg J.M., \& Taylor J.H., 2002, ApJ, 576, 942\\
%\bibitem{waaa} Weltevrede P., Abdo A. A., Ackerman M., Ajello M., Axelsson M., Baldini
% L., et al.~2010, ApJ, 708, 1426
%\bibitem{wj08a} Weltevrede P., \& Johnston S., 2008a, MNRAS, 391, 1210
%\bibitem{wj08b} Weltevrede P., \& Johnston S., 2008b, MNRAS, 387, 1755
%\bibitem{ww09} Weltevrede P., \& Wright G.A.E., 2009, MNRAS, 395, 2117
%\bibitem{wes06} Weltevrede, P., Edwards, R.T., \& Stappers, B.W. 2006,
%       A\&A, 445, 243
%\bibitem{wes07} 
Weltevrede P., Stappers B.W., \& Edwards R.T. B.W., 2007,
       A\&A, 469, 607\\
%\bibitem{w03} 
Wright, G. A. E., 2003, MNRAS, 344, 1041\\
%\bibitem{w04} Wright, G. A. E., 2004, MNRAS, 351, 813
%\bibitem{xkj98} Xilouris K. M., Kramer M., Jessner A., von Hoensbroech
%A., Lorimer D., et al., 1998, ApJ, 501, 286
%\bibitem{xqh97} Xu, R.X., Qiao, G.J., \& Han, J.L. 1997, A\&A, 323, 395
%\bibitem{yms11} Yan W.M., Manchester R.N., van Straten W., 
%Reynolds J.E., Hobbs G., et al.~2011, MNRAS, 414, 2087
%\bibitem{yhc2007}
You, X.P., Hobbs, G., Coles, W. A., Manchester, R. N.,
Edwards, R., Bailes, M.; Sarkissian, J., Verbiest, J. P. W.,
MNRAS, 378, 493\\
%\bibitem{zcwl} 
Zepka,A., Cordes, J. M., Wasserman, I., \& Lundgren, S. C. 1996, ApJ,
456, 305
%\bibitem{zs79} Zheleznyakov, V. V., \& Shaposhnikov, V.E. 1979, Aust. J. Phys., 32, 49
%\end{thebibliography}

\appendix
\section{Centroid shift for elongated subbeams}

\begin{figure}
	\centering
	\includegraphics[width=0.48\textwidth,]{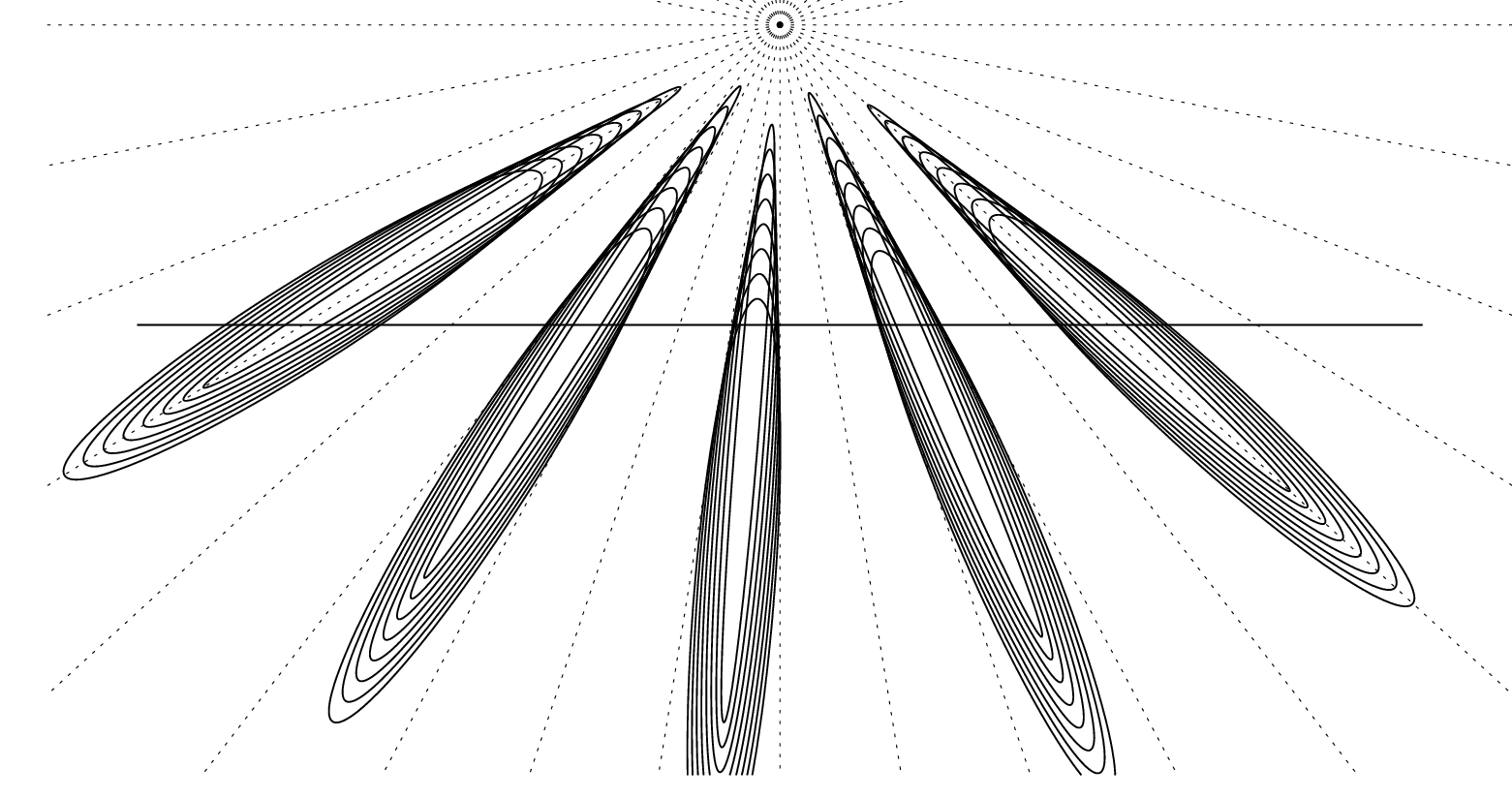}
\caption{A system of beams misaligned with respect to the MM.
The fan beams are elongated and spectrally nearly-uniform
(a case with little or no apparent RFM). 
The horizontal marks the passage of the line of
sight. Unlike in the case of the circular fixed-$\nu$ patterns
(cf.~Figs.~\ref{centroid} and \ref{misal}) the lag of the central component
appears for the leftward misalignment of the system.
}
\label{deromili}
\end{figure}

In the case of the elongated and 
broad-band beams there is little or no apparent 
RFM, as illustrated in Fig.~\ref{specgrad}c.
Longitudes of components are then determined by the cross-section of 
the sightline path with the azimuth of the fan beams:
\begin{equation}
\phi = \beta \tan\phm
\label{fabta}
\end{equation}
where $\beta$ is the impact angle (Fig.~\ref{params}).  
In contrast to the circular case discussed in Sect.~\ref{mathacc} (see
eq.~\ref{faza}), the observed pulse phase is now increasing with $\phm$ at 
a faster-than-linear rate of $\tan\phm$.
Therefore,
the LS spaces
between adjacent components are larger than the `corresponding' 
TS spaces,
when the system is rotated leftward (Fig.~\ref{deromili}), 
ie.~in the opposite direction than in
Fig.~\ref{centroid}.

To calculate the centroid shift, we use
the same intra-system magnetic azimuth $\Phi$ as defined in
Sect.~\ref{mathacc}, ie.~$\phm=\Phi+\epsilon$, where the misalignment angle
$\epsilon$ is now negative. 
The longitudes of the LS and TS
components then become: $\phi_{\sss L}=\beta\tan(-|\Phi_{\sss L}|-|\epsilon|)$
and $\phi_{\sss T}=\beta\tan(\Phi_{\sss T}-|\epsilon|)$, respectively. 
For a perfectly symmetric beam system $|\Phi_{\sss L}|=\Phi_{\sss T}=\Phi$.
The location of a pair centroid is 
\begin{equation}
\phi_{\rm pair} = \frac{\phi_{\sss L}+\phi_{\sss T}}{2} =
\frac{\beta\tan\epsilon\ (1 +\tan^2\Phi)}{1 - \tan^2\Phi\tan^2\epsilon}.
\label{centroelon}
\end{equation}
Note that $\phi_{\rm pair} < 0$ since $\epsilon < 0$ for the misalignment
shown in Fig.~\ref{deromili}. 
The central component (`core') is located at the (negative) pulse phase
\begin{equation}
\phi_{\rm ctc}=\beta\tan\epsilon
\label{coreposelon}
\end{equation}
and lags the centroid by the pulse phase interval
\begin{equation}
%\phi_{\rm pair} - \phi_{\rm ctc} = -\beta\tan\epsilon \ 
%\frac{\tan^2\Phi(\tan^2\epsilon + 1)}{1 - \tan^2\Phi\tan^2\epsilon}.
\Delta\phi_{\rm obs} = \phi_{\rm pair} - \phi_{\rm ctc} = 
\frac{\beta\tan\epsilon\tan^2\Phi(\tan^2\epsilon + 1)}
{1 - \tan^2\Phi\tan^2\epsilon}.
\label{lagelon}
\end{equation}
The value of $\Delta\phi_{\rm obs}$ is negative, because $|\phi_{\rm pair}| > 
|\phi_{\rm ctc}|$ and $\epsilon < 0$.
In the limit of small $|\epsilon|$ (and $\Phi$ not too close to 
$90^\circ$) the lag is equal to
\begin{equation}
%\phi_{\rm pair} - \phi_{\rm ctc} = -\beta\tan\epsilon \ 
%\frac{\tan^2\Phi(\tan^2\epsilon + 1)}{1 - \tan^2\Phi\tan^2\epsilon}.
\Delta\phi_{\rm obs} \approx \beta\tan\epsilon\tan^2\Phi
\label{lagelon2}
\end{equation}
For small $\Phi$ (inner components located close to the central one)
the centroid's location stays close to the position
$\phi_{\rm ctc}$ of the central component.
For increasing $\Phi$ (outer components) % at a large $\Phi$)
$\phi_{\rm pair}$ precedes the central component by a quickly
 increasing phase interval.
The magnitude of the `core' lag is then larger for the more peripheric
pairs, in agreement with the observations (GG01, GG03).
The effect increases with the misalignment $|\epsilon|$ of the beam system
(eq.~\ref{lagelon2}). 

Thus, for a bunch of similar, elongated 
 beams with a broadband spectrum, 
the misalignment shown in Fig.~\ref{deromili}
%the effect of different $\dcut$ alone 
can produce the forward shift of centroids
%, provided the beams are located slightly asymmetrically
with respect to the MM. Moreover, the shift is larger for the outer pairs 
of components. 
However, in this case of the weak spectral variations, 
the lag should exhibit little or no dependence on frequency,
and there should be no apparent RFM. The sign of the centroid shift 
would change (with the centroid following the central component) 
for the opposite misalignment than that of Fig.~\ref{deromili}.

\label{lastpage}
\end{document}